\documentstyle[preprint,aps,floats,epsf]{revtex}

\tightenlines

\begin{document}

\newcommand{\notE}{\ \hbox{{$E$}\kern-.60em\hbox{/}}}
\newcommand{\notp}{\ \hbox{{$p$}\kern-.43em\hbox{/}}}


\preprint{
\font\fortssbx=cmssbx10 scaled \magstep2
\hbox to \hsize{
\hfill$\vcenter{\hbox{\bf PSI-PR-99-02}
                \hbox{\bf MADPH-98-1091}
                \hbox{\bf hep-ph/9902202}
                \hbox{February 1999}}$ }
}
\title{\vspace*{2cm}
Parity Violating Asymmetries in Top Pair Production \\
at Hadron Colliders}
 
\author{Chung Kao$^a$\footnote{
Electronic Address: Kao@Pheno.physics.wisc.edu}
and D.~Wackeroth$^b$\footnote{
Electronic Address: Doreen.Wackeroth@psi.ch} }
\address{\vspace*{0.5cm}
$^a$ Department of Physics, University of Wisconsin, Madison,
Wisconsin 53706, USA\\
\vspace{0.2cm}
$^b$ Paul Scherrer Institut (PSI),
CH-5232 Villigen PSI, Switzerland}

\maketitle

\thispagestyle{empty}
 
\begin{abstract}

We study loop-induced parity violating asymmetries in the strong
production of polarized top quark pairs at $pp$ and $p \bar p$
colliders.  The electroweak ${\cal O}(\alpha)$ corrections to the
helicity amplitudes of $q \bar q \to t \bar t$ and $gg \to t \bar t$
are evaluated in a two Higgs doublet model (2HDM) and the minimal
supersymmetric standard model (MSSM).
While observables in top quark pair production receive little
contribution from standard electroweak interactions, it is possible
that they can be significantly enhanced in a 2HDM and the MSSM.
We find that the one-loop MSSM electroweak corrections can generate
parity violating asymmetries in the total production rate of left- and
right-handed top quark pairs up to about $1.7\%$ at the upgraded
Tevatron ($\sqrt{S}=2$ TeV) and $3\%$ at the LHC ($\sqrt{S}=14$ TeV).

\end{abstract}

%


\section{Introduction}

In the near future, the upgraded Fermilab Tevatron $p \bar p$ collider
will be able to produce about $7 \times 10^4$ top quark pairs
($t\bar{t}$) \cite{topacc} and the CERN Large Hadron Collider (LHC)
$pp$ collider will be able to generate about $10^7$ $t\bar{t}$ with a
moderate integrated luminosity of ${\cal L} = 10$ fb$^{-1}$.  The
large amount of top quark pairs suggests that it might be possible to
explore top quark observables and to perform electroweak precision
physics studies in the spirit of the successful LEP/SLD program
\cite{ewwg}.

Although the Standard Model (SM) has impressive experimental success,
various theoretical arguments such as the fine tuning problem and the
hierarchy problem suggest that the SM is merely the effective low
energy version of a more fundamental theory.  One attractive extension
of the SM is supersymmetry (SUSY) \cite{susy,MSSM} which connects
fermions and bosons.

We consider a two Higgs doublet model (2HDM) \cite{Guide} with two
doublets $\phi_1$ and $\phi_2$ that couple to the $t_3 = -1/2$ and
$t_3 = +1/2$ fermions, respectively.  After spontaneous symmetry
breaking, there remain five physical Higgs bosons: a pair of singly
charged Higgs bosons $H^{\pm}$, two neutral CP-even scalars $H^0$
(heavier) and $h^0$ (lighter), and a neutral CP-odd pseudoscalar
$A^0$.  The Higgs potential as well as the Yukawa interactions between
fermions and Higgs bosons in this 2HDM are the same as those of the
minimal supersymmetric standard model (MSSM) \cite{MSSM,Guide}.

The Higgs sector of a supersymmetric theory must contain at least two
$SU(2)$ doublets for anomaly cancellation.  In the MSSM, the Higgs
potential is constrained by supersymmetry such that all tree-level
Higgs boson masses and couplings are determined by just two
independent parameters, commonly chosen to be the mass of the CP-odd
pseudoscalar ($M_A$) and the ratio of vacuum expectation values of
Higgs fields ($\tan\beta \equiv v_2/v_1$).  If all supersymmetric
particles are much heavier than SM particles and Higgs bosons, the
MSSM becomes similar to a two Higgs doublet model (2HDM).

At the Tevatron, quark antiquark annihilation ($q\bar q \to t\bar t$)
is the major source of top quark pair production.  Gluon fusion ($gg
\to t\bar t$) produces most of the top quark pairs at the LHC.
Complementary to the direct search for signals of new physics, the
confrontation of theoretical predictions of top quark observables
beyond leading order in perturbation theory with their precise
measurements not only provides a consistency check for the SM but also
might reveal the nature of new physics.
The SM radiative corrections to both production mechanisms have been
calculated with one-loop QCD corrections as well as gluon resummation
\cite{qcd} and the electroweak (EW) ${\cal O}(\alpha)$ contribution.
The SM EW corrections have only marginal effects on $t \bar t$
observables \cite{diplpubsm,ttsm} such as the total cross section, the
invariant mass distribution and the asymmetries in the production of
left- and right-handed top quark pairs.  It is possible that these
observables might be considerably enhanced in supersymmetric models
and two Higgs doublet models.

Recent studies on radiative corrections to $q\bar q \to t\bar t$ and
$gg \to t\bar t$ in a general 2HDM and the MSSM found promisingly
large effects.  The one-loop EW corrections within the 2HDM
\cite{topthdm,topmssm} typically reduce the cross sections by up to
$\approx 6 \%$ at the upgraded Tevatron and $\approx 9\%$ at the LHC.
In most of the MSSM parameter space, the one-loop EW corrections
\cite{topmssm,sewk,chung,qcdew} reduce the cross sections by up to
$\approx 5,10 \%$ (Tevatron, LHC).

The SUSY QCD ${\cal O}(\alpha_s)$ corrections can enhance the effects
of the MSSM EW one-loop corrections depending on the choice of the
input parameters \cite{qcdew,sqcd1,sqcd2}.  The SUSY QCD corrections
to the $q \bar q$ annihilation subprocess, for instance, can
considerably affect the cross section when the gluino is not too
heavy, $m_{\tilde g} \alt 350$ GeV, so that at the upgraded Tevatron,
the combined MSSM EW and SUSY QCD one-loop corrections can reduce the
$t \bar t$ production cross section by up to $ \approx 10 \% $
($m_{\tilde g}=150$ GeV) \cite{sqcd2}.

In this paper we study parity violating asymmetries in polarized $t
\bar t$ production at hadron colliders.  We calculate the polarized
differential cross sections to ${\cal O}(\alpha \alpha_s^2)$ for both
$q\bar q \to t\bar t$ and $gg \to t\bar t$ within the 2HDM and the
MSSM and study the resulting asymmetries in the total production rate
and invariant mass distribution of left- and right-handed top quark
pairs at the upgraded Tevatron and the LHC. These
polarization asymmetries directly probe the parity non-conserving part
of the EW interactions within the models under consideration.  While
observables in unpolarized $t \bar t$ production have the drawback
that loop-induced effects need to be disentangled from the SM QCD
background, parity violating asymmetries in polarized strong $t \bar
t$ production have the potential to provide a clean signal of new
physics: there are no parity violating asymmetries at leading order in
perturbation theory since QCD preserves parity and the SM EW induced
asymmetries are too small to be observable, at least at the Tevatron
$p \bar p$ collider.  Thus, any large signal of parity non-conserving
effects in strong $t \bar t$ production suggests the presence of new
physics.  Within the models under consideration there is the
possibility of considerable enhancements of parity non-conserving EW
interactions due to the virtual presence of a charged Higgs boson with
enhanced Yukawa couplings, and of neutralinos and charginos, {\em
i.e.} the linear combinations of the supersymmetric partners of the
Higgs bosons and the electroweak gauge bosons.
%
%

The information on the polarization of the top quark can be deduced
from its decay products, since in average the top quark decays before
it can form a hadronic bound state or flips its spin.  How to measure
polarization asymmetries in $t\bar t$ production at hadron colliders
has been discussed in the context of testing CP violating
effects~\cite{cpviol} and studying spin correlations between the $t$
and $\bar t$ \cite{spincorr,stephen,yael}.  The polarization of the
(anti)top quark is transferred to its decay products, so that spin
information can be obtained from the resulting angular correlations
between the $t$ and $\bar t$ decay products or from the angular and/or
energy distribution of the charged lepton $l$ in $t\to b l\nu$, for
instance. In this paper we concentrate on asymmetries in the invariant
mass distribution and the total production rate of left- and
right-handed top quark pairs.  We will take into account the (anti)top
quark decay and study how the parity violating effects in the $t \bar
t$ production process affect observables involving the $t(\bar t)$
decay products in a forthcoming publication.


Recent measurements of the $b \to s\gamma$ decay rate by the CLEO
\cite{CLEO} and LEP collaborations \cite{LEP} place constraints on the
mass of the charged Higgs boson in a 2HDM with Yukawa interactions of
model II\footnote{\vspace*{0.5cm} The upper limit of $B(b\to s\gamma)$
demands that ($M_{H^\pm} \agt 240$ GeV) \cite{CLEO}.}  and on the
parameter space of the MSSM and the minimal supergravity unified model
(mSUGRA) \cite{bsgSUSY,Baer}.  
It was found that the branching ratio of $b \to s\gamma$ 
disfavors a large region of the MSSM and the mSUGRA parameter space
when $\tan\beta$ is large ($\tan\beta \agt 10$) and $\mu < 0$ \cite{Baer}.  
Therefore, we consider $\mu > 0$ in our
analysis\footnote{ In our convention, $+\mu$ appears in the chargino
mass matrix and $-\mu$ appears in the neutralino mass matrix.}.

This paper is organized as follows: \\ 
In Section II we present the
differential cross sections to ${\cal O}(\alpha \alpha_s^2)$ for
polarized strong $t\bar t$ production at hadron colliders within the
SM, the 2HDM and the MSSM.  In Section III we introduce the
polarization asymmetries in the polarized total $t \bar t$ production
rate and the invariant $t \bar t$ mass distribution which result from
the presence of parity violating EW interactions within the 2HDM and
the MSSM. We discuss the dependence on the input parameters of the
models under consideration with special emphasis on the numerical
significance compared to parity violating effects within the SM. We
conclude with Section IV.  The appendices provide explicit expressions
for the helicity amplitudes at the Born-level (Appendix A), for the
form factors parameterizing the parity non-conserving effects (Appendix
B), and for the interference between the one-loop EW corrections and
the Born matrix elements in the helicity states of $t\bar{t}$
(Appendix C).

\section{Polarized strong top pair production}

The main production mechanism for $t\bar t$ production at the Tevatron
is the annihilation of a quark-antiquark pair
\[ q(p_4) + \overline{q}(p_3) \rightarrow 
t(p_2) + \overline{t}(p_1) \]
whereas at the LHC the top quark pairs are mainly produced
via the fusion of two gluons
\[ g(p_4) + g(p_3) \rightarrow t(p_2) + \overline{t}(p_1) \; .\]
At the parton level, we obtain the corresponding differential cross
sections to polarized top pair production by applying spin projection
operators
\begin{eqnarray}\label{eq:one}
u_t(p_2,\lambda_t) \bar u_t(p_2,\lambda_t)& = & (1+2\lambda_t
\gamma_5 \not\!{s}_t) 
\frac{(\not\!{p}_2+m_t)}{2} \nonumber\\
v_{\bar t}(p_1,\lambda_{\bar t}) \bar v_{\bar t}(p_1,\lambda_{\bar t}) 
& = & (1+2\lambda_{\bar t} \gamma_5 
\not\!{s}_{\bar t})\frac{(\not\!{p}_1-m_t)}{2}
\end{eqnarray}
when contracting the matrix elements $\delta {\cal M}_i, i=q\bar q,gg$
describing the ${\cal O}(\alpha)$ contribution with the Born matrix
elements ${\cal M}_B^i$
\begin{eqnarray}\label{eq:two}
\frac{d \hat \sigma_i
(\hat t,\hat s,\lambda_t,\lambda_{\bar t})}{d\cos\hat\theta}
&= & \frac{d \hat \sigma^i_B
(\hat t,\hat s,\lambda_t,\lambda_{\bar t})}{d \cos\hat \theta}
+ \delta \frac{d \hat \sigma_i
(\hat t,\hat s,\lambda_t,\lambda_{\bar t})}{d \cos\hat \theta}
\nonumber\\
&=& \frac{\beta_t}{32 \pi \hat s} \,
\left[\overline{\sum} \mid {\cal M}^i_B \mid^2+
2 {\cal R}e \; \overline{\sum}
(\delta {\cal M}_i \times {\cal M}_B^{i*})\right] 
+ {\cal O}(\alpha^2 \alpha_s^2)\; ,
\end{eqnarray}
where 
$\beta_t=\sqrt{1-\frac{4 m_t^2}{\hat s}}$ is the top quark velocity; 
$\lambda_t$ $(\lambda_{\bar t}) = \pm 1/2$ denotes the helicity states 
of the top (anti-top) quark; 
$\hat s=(p_3+p_4)^2$ and $\hat t=(p_2-p_4)^2$ are Mandelstam variables; and 
$\hat\theta$ denotes the scattering angle of the top quark 
in the parton center of mass system (CMS).  
In Eq.~(\ref{eq:two}) the matrix elements squared 
are averaged over initial state spin and color degrees of freedom 
and summed over final state color degrees of freedom.
The spin four-vectors $s_{t,\bar t}$ are defined in
Appendix A, where we also provide explicit expressions for the squared
Born matrix elements ${\cal M}_B^i$ to the production of polarized
$t\bar t$ pairs.  The next-to-leading order matrix elements $\delta
{\cal M}_{q\bar q}$ and $\delta {\cal M}_{gg}$ comprise the EW ${\cal
O}(\alpha)$ corrections to the $q\bar q$ annihilation and gluon fusion
subprocesses, respectively, within the model under consideration.  In
the following presentation of polarized strong $t\bar t$ production to
${\cal O}(\alpha \alpha_s^2)$, we closely follow the notations of
Refs.~\cite{diplpubsm} (SM) and \cite{topmssm} (2HDM and MSSM), where
the impact of EW one-loop contributions on the unpolarized $t\bar t$
production cross sections has been studied.  Here we concentrate on
parity violating effects and give explicit expressions only for the
parity non-conserving parts of the EW ${\cal O}(\alpha)$ corrections.

\subsection{The Amplitudes}

At ${\cal O}(\alpha)$ within the SM the $gt\bar t$ vertex is modified
through the virtual presence of the Higgs boson $\eta$ and the
electroweak gauge bosons $Z^0$ and $W^{\pm}$.  In a two Higgs doublet
model, the SM Higgs sector is extended to have two Higgs doublets so
that the $gt\bar t$ vertex is now modified by the exchange of a heavy
and a light neutral scalar, $H^0$ and $h^0$, a pseudoscalar $A^0$ and
a charged Higgs boson $H^{\pm}$.  Within the MSSM $\delta {\cal
M}_{q\bar q, gg}$ comprise the EW contributions of the MSSM Higgs
sector and the genuine SUSY contributions, {\em i.e.}, the
modification of the $gt\bar t$ vertex by the virtual presence of
charginos $\tilde \chi^{\pm}_i$ and sbottoms $\tilde b_{L,R}$ and
neutralinos $\tilde \chi^0_i$ and stops $\tilde t_{1,2}$.  Here we
label the stop mass eigenstates such that $m_{\tilde{t}_1} <
m_{\tilde{t}_2}$ which is opposite to the notation chosen in
Ref.~\cite{topmssm}.

The ${\cal O}(\alpha)$ corrections can be parameterized in terms of
form factors revealing the Lorentz structure of the matrix elements to
strong top pair production as follows:

\medskip 

\underline{$q\bar q$ annihilation:}
\[\delta {\cal M}_{q\bar q} = \delta {\cal M}_{q\bar q}^{in}+
\delta {\cal M}_{q\bar q}^{fin} \]
with 
\begin{eqnarray}\label{eq:three}
\delta {\cal M}_{q\bar q}^{in}
& = & \alpha \alpha_s  \frac{i T^c_{ik} T^c_{jl}}{\hat s} 
\bar{u}_t^j(p_2) \gamma_{\mu} v_{\bar t}^l(p_1) 
\, \bar{v}_{\bar q}^k(p_3)\gamma^{\mu} \,(F_V(\hat s,0)
+\gamma_5 \, G_A(\hat s,0)) \, u_{q}^i(p_4)
\\ \label{eq:four}
\delta {\cal M}_{q\bar q}^{fin}
& = & \alpha \alpha_s  \frac{i T^c_{ik} T^c_{jl}}{\hat s} 
\bar{u}_t^j(p_2) \Bigl[\gamma_{\mu} \, 
(F_V(\hat s,m_t) +\gamma_5 \, G_A(\hat s,m_t)) \Bigr.
\nonumber\\
&+& \Bigl. (p_1-p_2)_{\mu}\, \frac{1}{2 m_t} F_M(\hat s,m_t)  
\Bigr] v_{\bar t}^l(p_1) 
\, \bar{v}_{\bar q}^k(p_3)\gamma^{\mu} \, u_{q}^i(p_4) \; ,
\end{eqnarray}
where $i,j,k,l;c$ are color indices and $T^c=\lambda^c/2$ with the
Gell-Mann matrices $\lambda^c$. $\delta {\cal M}_{q\bar q}^{in}$
describes the modification of the initial state $g q\bar q$ vertex
with the initial state quarks considered to be massless.  The final
state EW one-loop contribution to the $q\bar q$ annihilation
subprocess is described by $\delta {\cal M}_{q\bar q}^{fin}$.  The
form factors $F_{V,M}$ parameterize the parity conserving part of the
EW one-loop corrections and can be found in Refs.~\cite{diplpubsm}
(SM) and \cite{topmssm} (2HDM and MSSM).  The parity violating form
factors $G_A$ are explicitly given in Appendix B.

\medskip 

\underline{gluon fusion:}\\

\noindent
The ${\cal O}(\alpha)$ contribution to the gluon fusion subprocess
comprises the vertex corrections to the $s$ and $t(u)$ production
channel described by $F_{V,M},G_A$ and
$\rho_i^{V,t(u)},\sigma_i^{V,t(u)}$, respectively, the off-shell top
self energy insertion $\rho_i^{\Sigma,(t,u)},
\sigma_i^{\Sigma,(t,u)}$, the box diagrams $\rho_i^{\Box,(t,u)},
\sigma_i^{\Box,(t,u)}$ and the $s$-channel Higgs-exchange diagrams
$\rho_{12}^{\triangleleft}$
\begin{eqnarray}\label{eq:five}
\delta {\cal M}_{gg} &  = &  \alpha \, \alpha_s \, \left\{
\frac{f_{abc} T^c_{jl}}{\hat s} \, 
\left[ (M_2^{V,t}-2 M_3^{V,t}) \, F_V(\hat s,m_t)+
(M_2^{A,t}-2 M_3^{A,t}) \, G_A(\hat s,m_t)
\right. \right.
\nonumber\\
&+& \left. \left. 
( (\hat t-\hat u) \, M_{12}^{V,t}-4 M_{15}^{V,t}+4 M_{17}^{V,t} )
\, \frac{F_M(\hat s,m_t)}{2 m_t^2}\right] \right.
\nonumber\\
&+ & \left. \sum_{i=1,\ldots,7 \atop 11,\ldots,17}
\Bigl( i \,T^a_{jm} T^b_{ml} \, M_i^{V,t} \,\left[
\frac{\rho_i^{V,t}(\hat t,\hat s)}{\hat t-m_t^2}
+\frac{\rho_i^{\Sigma,t}(\hat t,\hat s)}{(\hat t-m_t^2)^2}
+\rho_i^{\Box,t}(\hat t,\hat s) \right] 
+ i \,T^b_{jm} T^a_{ml} \, M_i^{V,u} \, 
[t\rightarrow u] \Bigr. \right.  
\nonumber\\
&+& \left. \Bigl. i \,T^a_{jm} T^b_{ml} \, M_i^{A,t} \,
\left[\frac{\sigma_i^{V,t}(\hat t,\hat s)}{\hat t-m_t^2}
+\frac{\sigma_i^{\Sigma,t}(\hat t,\hat s)}{(\hat t-m_t^2)^2}
+\sigma_i^{\Box,t}(\hat t,\hat s) \right]
+ i \, T^b_{jm} T^a_{ml} \, M_i^{A,u} \, [t\rightarrow u] \Bigr)
\right.
\nonumber\\
&+&\left.  \sum_{S=\eta \atop S=H^0,h^0} \frac{(-i \, \delta_{jl} 
\delta^{ab}) \, M_{12}^{V,t} \,
\rho_{12}^{\triangleleft}(\hat s,M_S)}
{(\hat s-M_S^2)^2+ (M_S \,\Gamma_S)^2} \right\} \;.
\end{eqnarray}
The standard matrix elements $M_i^{(V,A),(t,u)}$ and the parity
conserving form factors $\rho_i^{X,(t,u)}$ are explicitly given in
Refs.~\cite{diplpubsm} (SM) and~\cite{topmssm} (2HDM and MSSM). In
Appendix B we provide explicit expressions for $\sigma_i^{X,(t,u)}$
which parameterize the parity violating part of the EW one-loop
corrections to the gluon fusion subprocess.

After contracting the next-to-leading order matrix elements $\delta
{\cal M}_{q\bar q,gg}$ with the Born matrix elements ${\cal
M}_B^{q\bar q,gg}$ according to Eq.~(\ref{eq:two}) the polarized
differential cross sections are described in terms of scalar products
involving the external four-momenta and the top/antitop spin
four-vectors $s_{t,\bar t}$.  The latter are defined after choosing
the axes along which the $t$ and $\bar t$ spins are decomposed as it
is described in Appendix A.  As studied in Refs.~\cite{stephen,yael}
the freedom in the choice of the spin axes can be used to increase
spin correlations at hadron colliders.  Here we choose the helicity
basis where the spin is quantized along the particle's direction of
motion.  In a forthcoming publication, once we have taken into account
the decay of the polarized top quarks we will also look into the
possibility to increase the observability of parity violating effects
by choosing other bases than the helicity basis.

\subsection{The Cross Sections}

Using the helicity basis the polarized differential $t\bar t$
production cross sections to ${\cal O}(\alpha \alpha_s^2)$ at the
parton level read as follows:\\

\underline{$q\bar q$ annihilation:}
\[\delta \,\frac{d \hat{\sigma}_{q\bar q}
(\hat s,\hat t,\lambda_t,\lambda_{\bar t})}{d \cos\hat \theta}
= \delta \,\frac{d \hat{\sigma}_{q\bar q}^{in}}{d \cos\hat \theta}+
\delta \,\frac{d \hat{\sigma}_{q\bar q}^{fin}}{d \cos\hat \theta}\]
with
\begin{eqnarray}\label{eq:six}
\delta \,\frac{d \hat{\sigma}_{q\bar q}^{in}
(\hat s,\hat t,\lambda_t,\lambda_{\bar t})}{d \cos\hat \theta}
& = & \frac{\alpha \, \alpha_s^2}{8\,  \hat s} 
\, \langle \frac{2}{9}\rangle \, \beta_t \, \frac{1}{4} 
\nonumber\\
&\times& 2 \, {\cal R}e \left\{   
[2-\beta_t^2 (1-\cos^2\hat \theta)-4 \lambda_t\lambda_{\bar t} 
(\beta_t^2 (1-\cos^2\hat \theta)
+2 \cos^2 \hat \theta) ]\, F_V(\hat s,0)
\right. \nonumber\\
&+& \left. 4 \cos\hat \theta \,  (\lambda_t-\lambda_{\bar t})\, 
G_A(\hat s,0)\right\}
\end{eqnarray}

\begin{eqnarray}\label{eq:seven}
\delta \,\frac{d \hat{\sigma}_{q\bar q}^{fin}
(\hat s,\hat t,\lambda_t,\lambda_{\bar t})}{d \cos\hat \theta}
& = & \frac{\alpha \, \alpha_s^2}{8\,  \hat s} 
\, \langle \frac{2}{9}\rangle \, \beta_t \, \frac{1}{4}
\nonumber\\
&\times& 2 \, {\cal R}e \left\{  
[2-\beta_t^2 (1-\cos^2\hat \theta)-4 \lambda_t\lambda_{\bar t} 
(\beta_t^2 (1-\cos^2\hat \theta)+2 \cos^2 \hat \theta)] 
\, F_V(\hat s,m_t)
\right. \nonumber\\
&+& \left. \beta_t^2 \, (1-\cos^2\hat \theta)\,
(1+4 \lambda_t\lambda_{\bar t})\, F_M(\hat s,m_t) \right.
\nonumber\\
&+& 2 \left. \beta_t \, (1+\cos^2\hat \theta)\,  
(\lambda_t-\lambda_{\bar t})\, G_A(\hat s,m_t)\right\}
\end{eqnarray}

\medskip

\underline{gluon fusion:}
\begin{eqnarray}\label{eq:eight}
\delta \,\frac{d \hat{\sigma}_{gg}
(\hat s,\hat t,\lambda_t,\lambda_{\bar t})}{d \cos\hat \theta} & = &
\frac{\alpha \alpha_s^2}{8 \, \hat s} \, \frac{1}{64} \, \beta_t 
\nonumber\\
& \times & 2\, {\cal R}e \left\{\sum_{j=1,2,3}
\left( c^s(j) \frac{1}{\hat s} 
\left[ M^{V,t}_{\lambda_t\lambda_{\bar t}}(2,j) \, F_V(\hat s,m_t)
\right. \right.\right.
\nonumber\\
&+& \left. \left. \left. 
M^{V,t}_{\lambda_t\lambda_{\bar t}}(12,j) \, 
\frac{(\hat t-\hat u)}{2m_t^2} F_M(\hat s,m_t)
+M^{A,t}_{\lambda_t\lambda_{\bar t}}(2,j) \, G_A(\hat s,m_t)
\right]\right.\right.
\nonumber\\
&+& \left. \left. c^t(j) \sum_{i=1,\ldots,7 \atop 11,\ldots,17} 
\left[ M^{V,t}_{\lambda_t\lambda_{\bar t}}(i,j) \, 
\left(\frac{\rho_i^{V,t}(\hat t,\hat s)}{\hat t-m_t^2}
+\frac{\rho_i^{\Sigma,t}(\hat t,\hat s)}{(\hat t-m_t^2)^2}
+\rho_i^{\Box,t}(\hat t,\hat s) \right) \right.\right.\right.
\nonumber \\
&+& \left.\left.\left.
M^{A,t}_{\lambda_t\lambda_{\bar t}}(i,j) \, 
\left(\frac{\sigma_i^{V,t}(\hat t,\hat s)}{\hat t-m_t^2}
+\frac{\sigma_i^{\Sigma,t}(\hat t,\hat s)}{(\hat t-m_t^2)^2}
+\sigma_i^{\Box,t}(\hat t,\hat s) \right) \right] \right.\right.
\nonumber\\
&+ & \left. \left. c^u(j) \sum_i 
\left[ M^{V,u}_{\lambda_t\lambda_{\bar t}}(i,j)
\,\left(\frac{\rho_i^{V,u}(\hat u,\hat s)}{\hat u-m_t^2}
+\frac{\rho_i^{\Sigma,u}(\hat u,\hat s)}{(\hat u-m_t^2)^2}
+\rho_i^{\Box,u}(\hat u,\hat s) \right) \right. \right. \right.
\nonumber\\
&+& \left.\left. \left.
M^{A,u}_{\lambda_t\lambda_{\bar t}}(i,j)
\,\left(\frac{\sigma_i^{V,u}(\hat u,\hat s)}{\hat u-m_t^2}
+\frac{\sigma_i^{\Sigma,u}(\hat u,\hat s)}{(\hat u-m_t^2)^2}
+\sigma_i^{\Box,u}(\hat u,\hat s)\right) \right] \right) \right.
\nonumber\\
&+ & \left. \sum_{S=\eta \atop S=H^0,h^0} 
\frac{4 \,(M^{V,t}_{\lambda_t\lambda_{\bar t}}(12,2)
+M^{V,t}_{\lambda_t\lambda_{\bar t}}(12,3))}{(\hat s-M_S^2)^2
+ (M_S \, \Gamma_S)^2} \, \rho_{12}^{\triangleleft}(\hat s,M_S)
\right\} \; ,
\end{eqnarray}
where $i$ numerates the 14 standard matrix elements
$M_i^{(V,A),(t,u)}$ and $j=1,2,3$ the $s,t,u$-channel of the Born
matrix element. The color factors $c^{s,t,u}(j)$ can be found in
Appendix B.3 of Ref.~\cite{diplpubsm}.  In Eq.~(\ref{eq:eight})
$M^{(V,A),(t,u)}_{\lambda_t\lambda_{\bar t}}(i,j)$ contains the
average over the gluon polarization states for fixed top/antitop
helicities according to Eq.~(3.40) of Ref.~\cite{diplpubsm}:
\begin{equation}\label{eq:nine}
M^{(V,A),(t,u)}_{\lambda_t\lambda_{\bar t}}(i,j) 
= \frac{1}{4}\sum_{gluon \atop polarization} 
M_i^{(V,A),(t,u)} \times M_B^{gg,j^*}  \; ,
\end{equation}
where $M_B^{gg,j}$ describes the $j$-th production channel
contribution to the Born matrix element.  In Appendix C we provide all
non-vanishing contractions of the standard matrix elements with the
Born matrix elements.

The observable hadronic cross sections to $p p,p \bar p \to t \bar t
X$ are obtained by convoluting the partonic cross sections of
Eqs.~(\ref{eq:six},\ref{eq:seven},\ref{eq:eight}) with the
corresponding parton distribution functions.  After performing the
integration over the scattering angle the $t\bar t$ observables of
interest in this paper, the polarized $t\bar t$ production cross
section $\sigma_{\lambda_t,\lambda_{\bar t}}$ and invariant $t\bar t$
mass distribution $d \sigma_{\lambda_t,\lambda_{\bar t}}/d M_{t\bar
t}$, are described as follows:
\begin{equation}\label{eq:ten}
\sigma_{\lambda_t,\lambda_{\bar t}}(S) 
= \int_{\frac{4 m_t^2}{S}}^1 \frac{d\tau}{\tau}
\left(\frac{1}{S} \frac{dL_{q\bar q}}{d\tau} \; 
\hat s \hat \sigma_{q\bar q}(\hat s,\alpha_s(\mu),
\lambda_t,\lambda_{\bar t})+\frac{1}{S} \frac{dL_{gg}}{d\tau} \; 
\hat s \hat \sigma_{gg}(\hat s,\alpha_s(\mu),
\lambda_t,\lambda_{\bar t})\right)
\end{equation}
with $\tau=x_1 x_2=\hat s/S$ and the parton luminosities
\begin{equation}\label{eq:eleven}
\frac{d L_{ij}}{d\tau}=\frac{1}{1+\delta_{ij}}
\int_{\tau}^1 \frac{dx_1}{x_1} \left[ f_i(x_1,Q)f_j(\frac{\tau}{x_1},Q)
+(1 \leftrightarrow 2)\right] \; ,
\end{equation}
and 
\begin{equation}\label{eq:twelve}
\frac{d \sigma_{\lambda_t,\lambda_{\bar t}}}{d M_{t\bar t}} 
= \sum_{ij=q\bar q,gg}\frac{2}{M_{t\bar t}}
\hat \sigma_{ij}(\hat s=\tau S,\lambda_t,\lambda_{\bar t})
\;  \tau \frac{d L_{ij}}{d \tau}
\end{equation}
with $\tau=M_{t\bar t}^2/S$.  We use the MRSA set of parton
distribution functions~\cite{mrsa} with the factorization ($Q$) and
renormalization scale ($\mu$) chosen to be $Q=\mu=m_t$.  In order to
take into account that jets originating from the produced top quarks
at large scattering angles are better distinguishable from the
background we impose a cut on the transverse momentum $p_t$ of the top
quark in the CM frame:~$p_t>$20, 100 GeV (Tevatron, LHC).  Finally, in
the following numerical evaluation the EW SM input parameters are
chosen to be~\cite{ewwg,ewpara}:
\[ m_t = 174 \, \mbox{GeV},\,  m_b=4.7 \, \mbox{GeV}, \, M_W = 80.39 \, 
\mbox{GeV}, \, M_Z= 91.1867 \, \mbox{GeV} \]
\[ \mbox{ and } \; \alpha^{-1}(M_Z) = 128.9 \; .\]

\section{Parity violating asymmetries}

In order to study the impact of loop-induced parity violating
interactions in strong $t\bar t$ production at the upgraded Tevatron
(with $\sqrt{S}=2$ TeV) and the LHC (with $\sqrt{S}=14$ TeV) we
introduce the following polarization asymmetries \cite{ttsm,chung}:
\begin{itemize}
\item
the differential left-right asymmetry 
\begin{equation}\label{eq:thirteen}
\delta {\cal A}_{LR}(M_{t\bar t}) 
= \frac{d\sigma_{{+\frac{1}{2},-\frac{1}{2}}}/ 
d M_{t\bar t}-d\sigma_{{-\frac{1}{2},+\frac{1}{2}}}
/ d M_{t\bar t}}{d\sigma_{{+\frac{1}{2},-\frac{1}{2}}}/ d M_{t\bar t}
+d\sigma_{{-\frac{1}{2},+\frac{1}{2}}}/ d M_{t\bar t}}
\end{equation}
with $d\sigma_{\lambda_t,\lambda_{\bar t}}/d M_{t\bar t}$ 
of Eq.~(\ref{eq:twelve}),
\item
the integrated left-right asymmetry
\begin{equation}\label{eq:fourteen}
{\cal A}_{LR} 
= \frac{\sigma_{{+\frac{1}{2},-\frac{1}{2}}}
-\sigma_{{-\frac{1}{2},+\frac{1}{2}}}}
{\sigma_{{+\frac{1}{2},-\frac{1}{2}}}
+\sigma_{{-\frac{1}{2},+\frac{1}{2}}}} 
\end{equation}
with $\sigma_{\lambda_t,\lambda_{\bar t}}(S)$ 
of Eq.~(\ref{eq:ten}) and
\item
the integrated left-right asymmetry when assuming that the
polarization of only one of the top quarks in $t \bar t$ is measured
in the experiment (here we sum over the $\bar t$ helicity states)
\begin{equation}\label{eq:fifteen}
{\cal A} = \frac{(\sigma_{{+\frac{1}{2},-\frac{1}{2}}}
+\sigma_{{+\frac{1}{2},+\frac{1}{2}}})
-(\sigma_{{-\frac{1}{2},+\frac{1}{2}}}
+\sigma_{{-\frac{1}{2},-\frac{1}{2}}})}{\sigma} \equiv 
\frac{\sigma_{{+\frac{1}{2},-\frac{1}{2}}}
-\sigma_{{-\frac{1}{2},+\frac{1}{2}}}}{\sigma}\; ,
\end{equation}
where $\sigma(S)=\sum_{\lambda_t,\lambda_{\bar t}=\pm 1/2} 
\sigma_{\lambda_t, \lambda_{\bar t}}$ 
is the total unpolarized $t\bar t$ production cross section.
\end{itemize}
When studying these asymmetries we are directly probing the parity
non-conserving interactions parameterized by the form factors $G_A$
and $\sigma_i^{X,(t,u)}$ of Eqs.~(\ref{eq:btwo},\ref{eq:bseven}).  For
illustration, we only consider the $q\bar q$ annihilation subprocess
so that the differential left-right asymmetry $\delta {\cal A}_{LR}$
of Eq.~(\ref{eq:thirteen}) can be approximated as follows:
\begin{equation}\label{eq:sixteen}
\delta {\cal A}_{LR}(M_{t\bar t}) 
\approx \frac{\alpha}{4\pi} \, \beta_t \, 2 {\cal R}e 
\, G_A(M_{t\bar t}) \; \Bigl(1+\frac{\alpha}{4\pi}
\, 2 {\cal R}e \, F_V(M_{t\bar t}) \Bigr)^{-1} \; ,
\end{equation}
which is a good approximation in case of $t\bar t$ production at the
Tevatron.  As can be seen the loop-induced left-right asymmetry is
directly proportional to the parity violating form factor $G_A$.

Before we start the numerical discussion we would like to point out
that there is another source of parity violation when the top quark
pairs are produced in the Drell-Yan process $q\bar q \rightarrow Z
\rightarrow t\bar t$ with the differential production cross section at
the parton level (using helicity basis)~\cite{chung}
\begin{eqnarray}\label{eq:seventeen}
\frac{d \hat \sigma_Z(\hat s,\lambda_t,\lambda_{\bar t})}
{d \cos\hat \theta} & = & 
\frac{\pi \alpha^2 \beta_t \hat s}{\mid\hat s-M_Z^2 \mid^2} 
\,  \left\{a_q v_q [(a_t^2 \beta_t^2+v_t^2) 
(\lambda_{\bar t}-\lambda_t)+(1-4\lambda_t\lambda_{\bar t}) 
a_t v_t \beta_t]\, \cos\hat \theta \right.
\nonumber\\
&+& \left. \frac{(a_q^2+v_q^2)}{8} \, 
[ \beta_t (1+\cos^2\hat \theta) 
(a_t^2 \beta_t (1-4 \lambda_t\lambda_{\bar t})
+4 (\lambda_{\bar t}-\lambda_t) a_t v_t)
\right. \nonumber\\
&+&\left.  v_t^2 (2-\beta_t^2 (1-\cos^2\hat \theta)
-4 \lambda_t \lambda_{\bar t}
(\beta_t^2 (1-\cos^2\hat \theta)+2 \cos^2\hat \theta)) ]\right\} \; ,
\end{eqnarray}
where $v_{q,t},a_{q,t}$ parameterize the $Z q\bar q, Z t \bar t$
couplings~\cite{diplpubsm}.  The Drell-Yan production process prefers
the production of $t_L \bar t_R$ pairs, so that the loop-induced
asymmetry in strong $t\bar t$ production at the Tevatron is modified
as follows:
\begin{equation}\label{eq:eighteen}
\delta {\cal A}_{LR}(M_{t\bar t}) 
\approx \beta_t \Bigl[\frac{\alpha}{4\pi} \, 2 {\cal R}e 
\, G_A(M_{t\bar t})-\Bigl(\frac{3 \,\alpha}{\alpha_s} 
\frac{M_{t\bar t}^2}{M_{t\bar t}^2-M_Z^2}\Bigr)^2 
a_t v_t (v_q^2+a_q^2)\Bigr]
\; \Bigl(1+\frac{\alpha}{4\pi}\, 2 {\cal R}e 
\, F_V(M_{t\bar t}) \Bigr)^{-1}   
\end{equation}
Thus, the parity violating asymmetries are either enhanced or reduced
depending on the model under consideration.  Within the SM, for
instance, the differential left-right asymmetry $\delta {\cal A}_{LR}$
in the production process $q\bar q\rightarrow t\bar t$ is less
pronounced when the Drell-Yan production mechanism is included.  In
the following numerical discussion, we always include the $q\bar q
\rightarrow Z \rightarrow t\bar t$ production process.  Naturally, at
the LHC the Drell-Yan production process is strongly suppressed and in
particular has no effect on the parity violating asymmetries.

\begin{figure}
\begin{center}
\setlength{\unitlength}{1cm}
\setlength{\fboxsep}{0cm}
\begin{picture}(16.4,18)
\put(-3,-6.5){\epsfbox{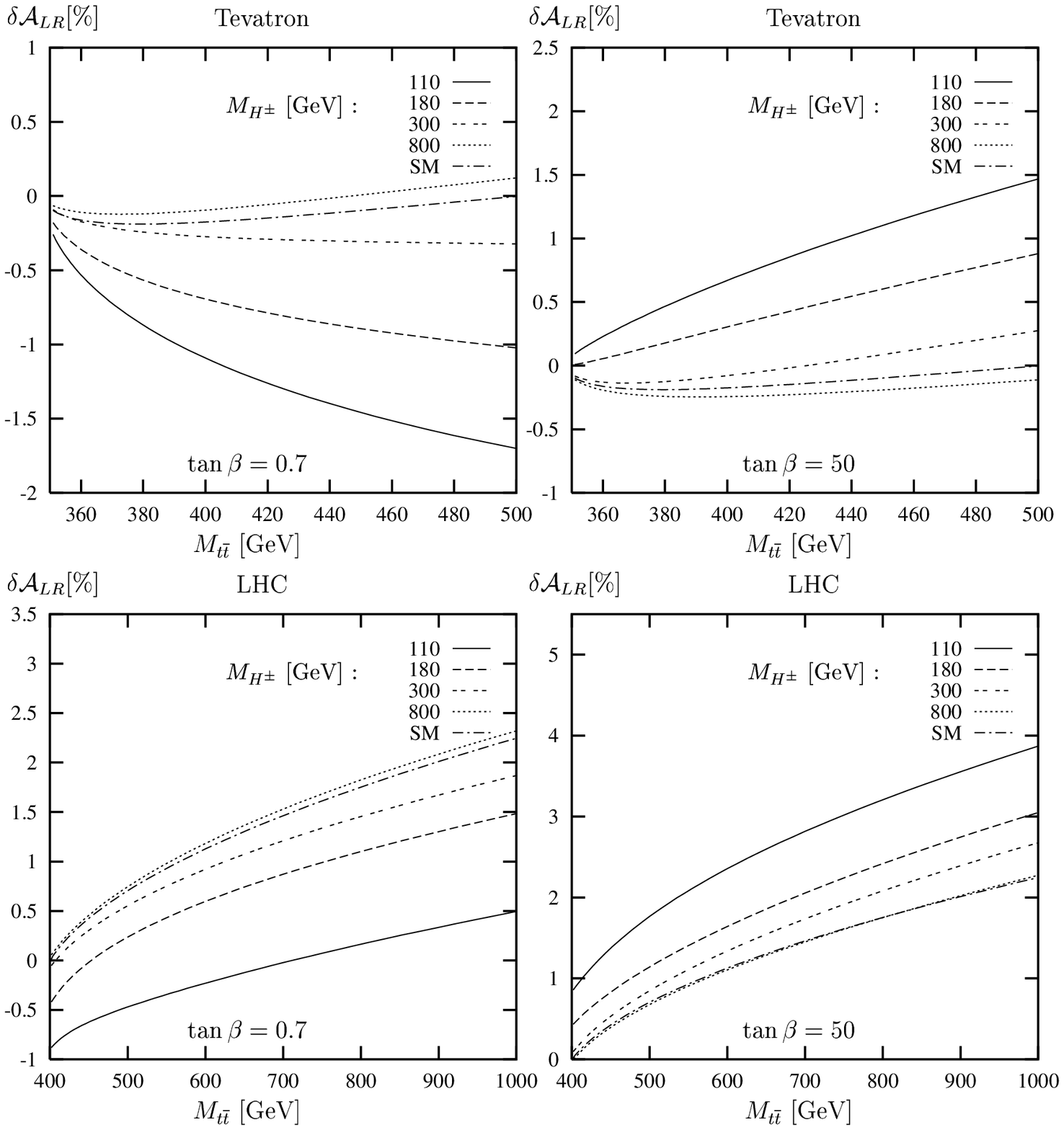}}
\end{picture}
\end{center}
\caption{The differential asymmetry $\delta {\cal A}_{LR}$ 
at the upgraded Tevatron and the LHC within the 2HDM for different
values of $M_{H^{\pm}}$ and $\tan\beta$ (with $\alpha=\pi/2$, $M_H=75$
GeV, $M_h=70$ GeV, $M_A=75$ GeV).}
\label{fig:one}
\end{figure}

\begin{figure}
\begin{center}
\setlength{\unitlength}{1cm}
\setlength{\fboxsep}{0cm}
\begin{picture}(16.4,18)
\put(-3,-6.5){\epsfbox{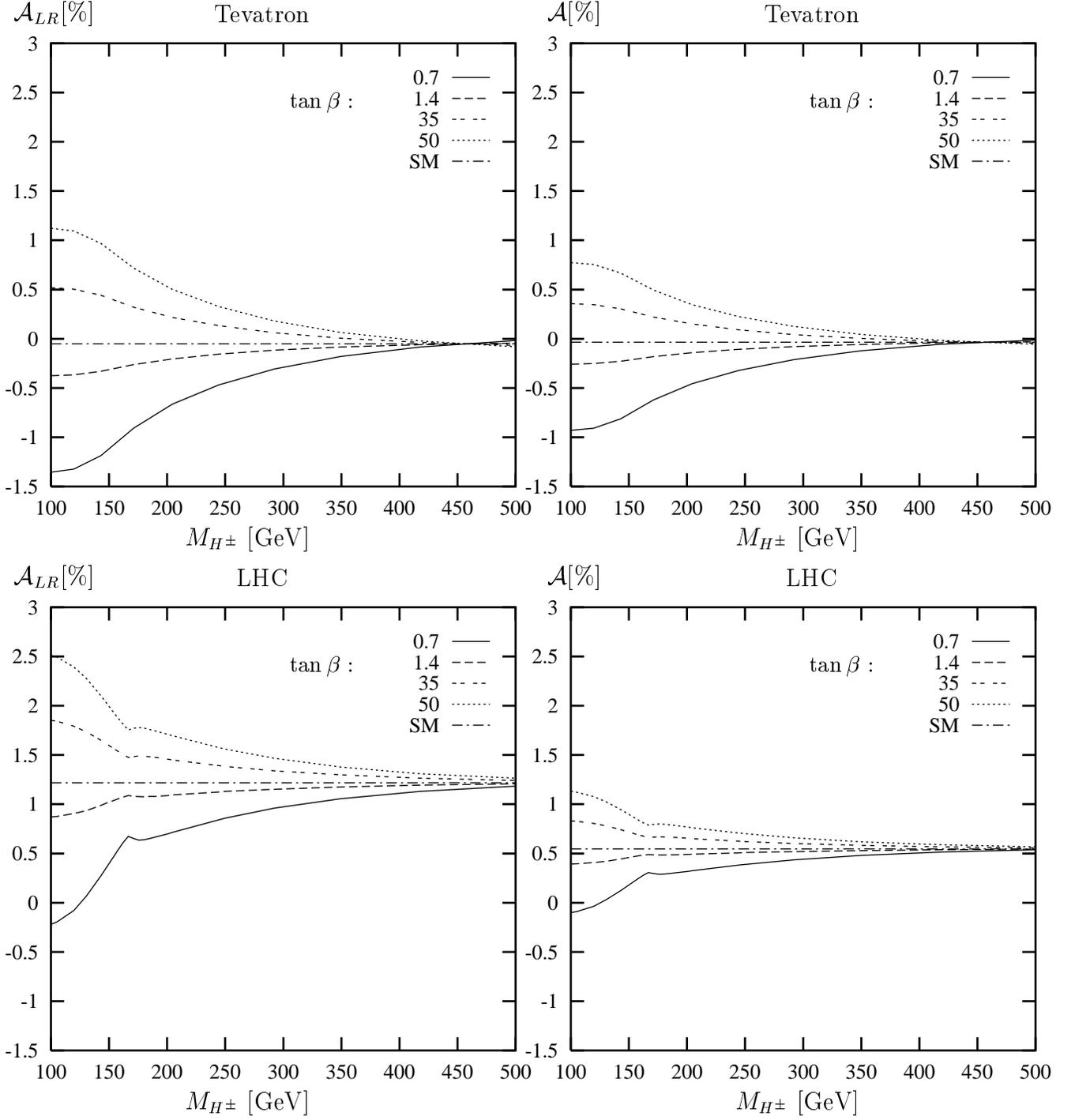}}
\end{picture}
\end{center}
\caption{The variation of the integrated asymmetries ${\cal A}_{LR}$ 
and ${\cal A}$ with $M_{H^{\pm}}$ at the upgraded Tevatron and the LHC
within the 2HDM for different values of $\tan\beta$ (with
$\alpha=\pi/2$, $M_H=75$ GeV, $M_h=70$ GeV, $M_A=75$ GeV).}
\label{fig:two}
\end{figure}

We start the numerical discussion with the study of loop-induced
parity violating effects within the 2HDM.  Within the 2HDM the
loop-induced asymmetry results from the virtual presence of
electroweak gauge bosons and the charged Higgs boson.  As can be
easily verified from the structure of the top-Yukawa coupling to the
charged Higgs boson (see, e.g., Table I of Ref.~\cite{topmssm}) there
are two possibilities of enhancement:~for very large and very small
values of $\tan\beta$.  This is illustrated in Fig.~\ref{fig:one}
where we compare $\delta {\cal A}_{LR}$ obtained within the 2HDM for
different choices of $\tan\beta$ and $M_{H^{\pm}}$ with the asymmetry
obtained within the SM (with $M_{\eta}=100$ GeV).  In
Fig.~\ref{fig:two} the integrated asymmetries ${\cal A}_{LR}$ of
Eq.~(\ref{eq:fourteen}) and ${\cal A}$ of Eq.~(\ref{eq:fifteen}), are
shown in dependence of $M_{H^{\pm}}$ for different values of
$\tan\beta$.  At the upgraded Tevatron and the LHC the parity
violating interactions within the 2HDM yield asymmetries in the total
production rate of left- and right-handed top quark pairs of $|{\cal
A}_{LR}| \le 1.4\%$ and $ \le 2.5\%$, respectively.  They can be
considerably larger than the asymmetries observed within the SM:
$|{\cal A}_{LR}|=0.05\%$ (Tevatron) and $1.2\%$ (LHC).  When we assume
that the polarization of the antitop quark is not measured the
remaining 2HDM asymmetry still amounts to $|{\cal A}| \le 0.9\%,
1.1\%$ compared to $|{\cal A}| = 0.04 \%, 0.5\%$ (Tevatron, LHC)
within the SM.

Within the MSSM additional sources of parity violation in strong
$t\bar t$ production occur due to genuine SUSY EW contributions, {\em
i.e.}  the parity violating components of the $t-\tilde t_j-\tilde
\chi^0_i$ and $t-\tilde b_j-\tilde \chi^{\pm}_i$ interactions.  For
the moment we shall neglect the parity violating effects induced by
SUSY QCD interactions, which only arise when the squarks are
non-degenerate in mass \cite{sqcd1,sqcd2}.  In the following we study
the dependence of the differential and integrated asymmetries on the
following MSSM input parameters:
\[\tan\beta, M_A, \mu, M_2, m_{\tilde b_L}, m_{\tilde t_1}, 
\Phi_{\tilde t} \; .\]
Alternatively, we can also choose $M_{H^{\pm}}$ as input parameter and
$M_A$ is determined through the relation $M_A^2=M_{H^{\pm}}^2-M_W^2$.
%
%
We evaluate the Higgs masses and couplings with one loop radiative
corrections \cite{Higgs,Bisset} from both the top and the bottom
Yukawa interactions at the scale $Q =
\sqrt{m_{\tilde{t}_L}m_{\tilde{t}_R}}$.  When this high scale is used,
the RGE improved one-loop corrections approximately reproduce the
dominant two loop perturbative calculation \cite{Two-loop} of the mass
of the lighter CP-even Higgs scalar ($M_h$).
To account for the experimental bounds on the supersymmetric particle
mass spectrum from the negative search at LEP we only allow those
parameter combinations which yield $M_h \ge 70$ GeV, 
the lightest neutralino $M_{\tilde\chi^0} \ge 25$ GeV 
and $M_{\tilde\chi^{\pm}} \ge 90$ GeV.

\begin{figure}
\begin{center}
\setlength{\unitlength}{1cm}
\setlength{\fboxsep}{0cm}
\begin{picture}(16.4,18)
\put(-3,-6.5){\epsfbox{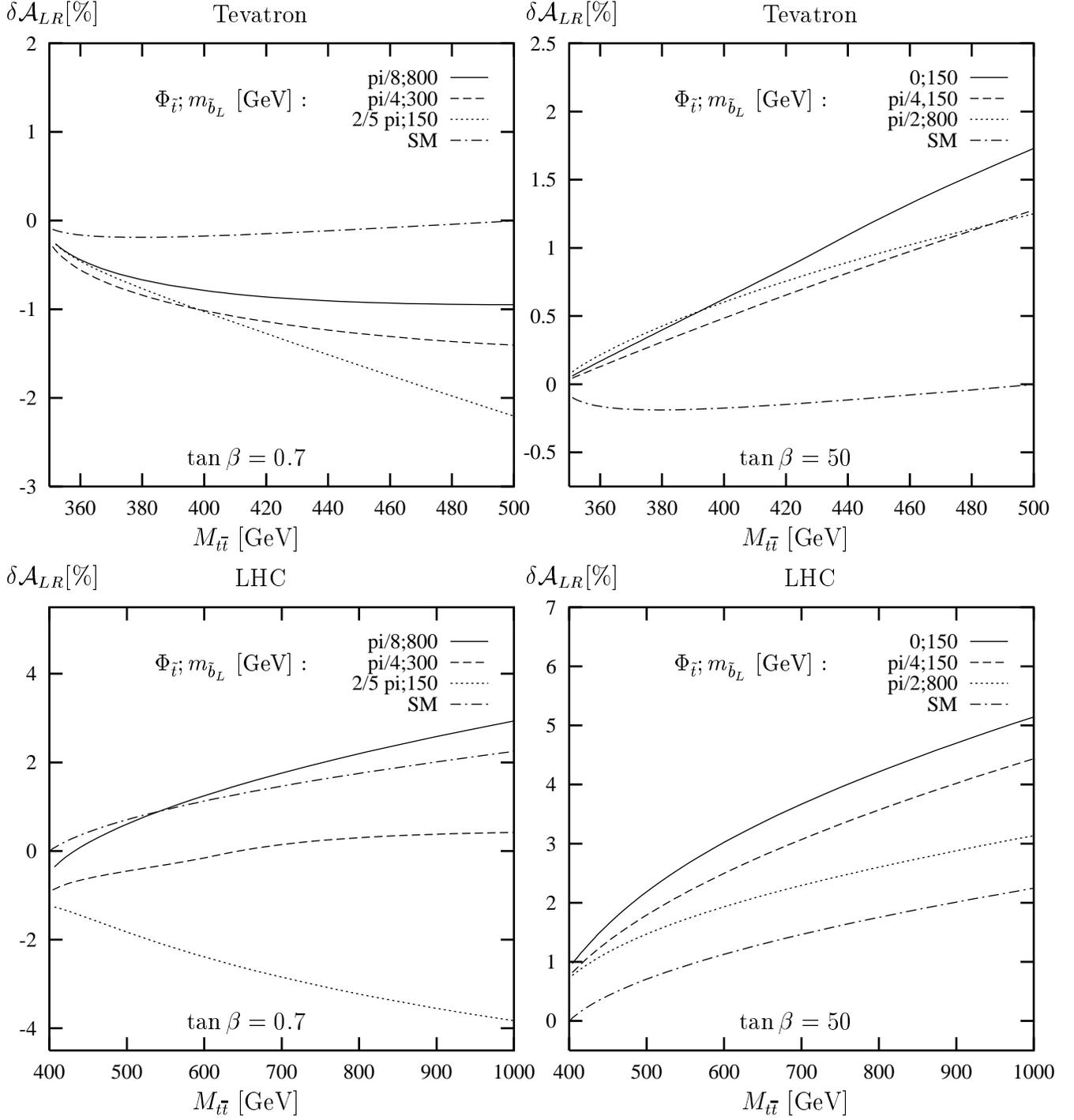}}
\end{picture}
\end{center}
\caption{The differential asymmetry $\delta {\cal A}_{LR}$ 
at the upgraded Tevatron and the LHC within the MSSM for 
$m_{\tilde t_1}=90$ GeV and
different values of $\Phi_{\tilde t}$ and
$m_{\tilde b_L}$ (with $M_{H^{\pm}}$=110 GeV, $\mu=120$ GeV and
$M_2=3|\mu|$).} 
\label{fig:three}
\end{figure}

\begin{figure}[htb]
\begin{center}
\setlength{\unitlength}{1cm}
\setlength{\fboxsep}{0cm}
\begin{picture}(16.4,18)
\put(-3,-6.5){\epsfbox{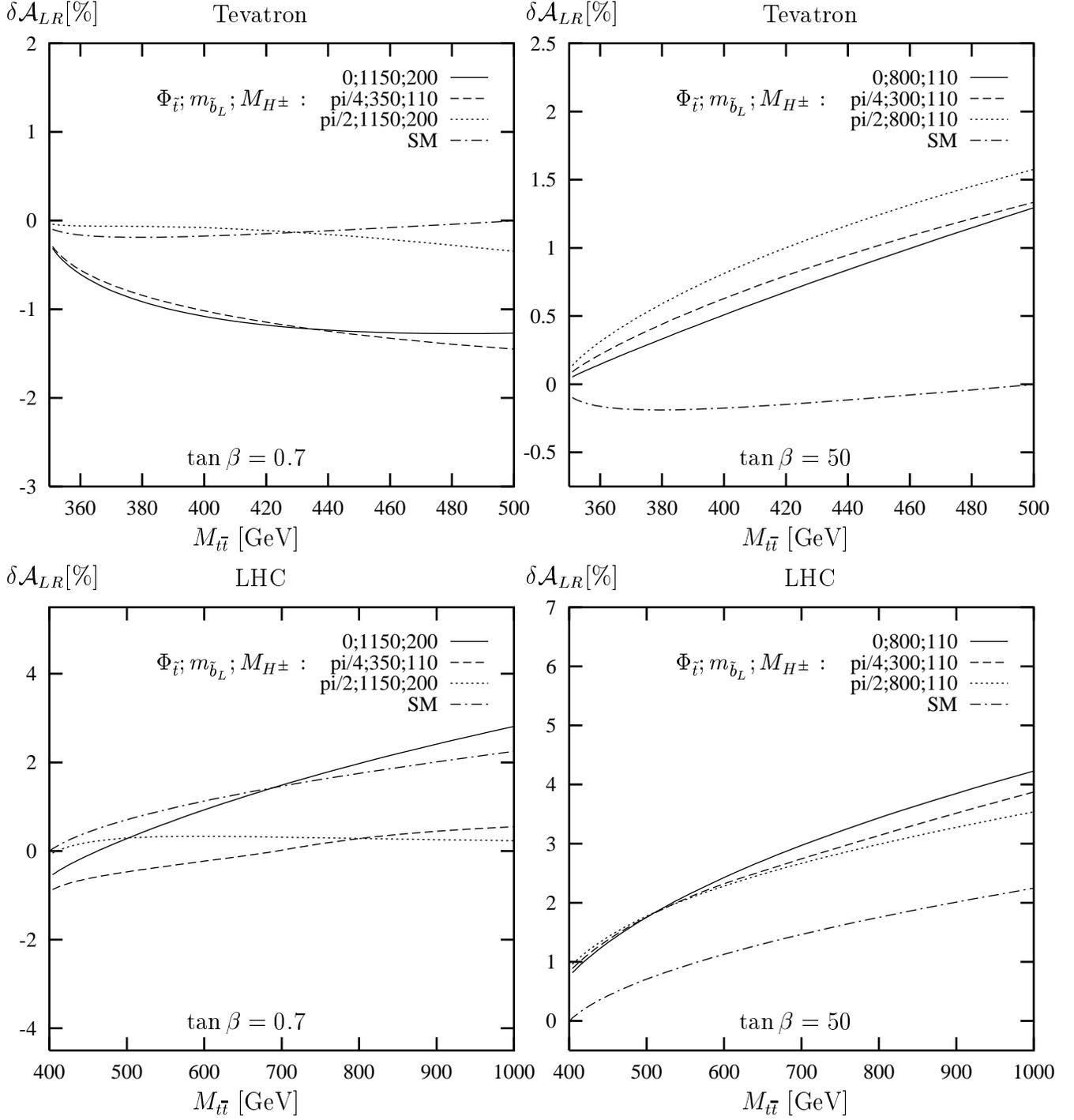}}
\end{picture}
\end{center}
\caption{The differential asymmetry $\delta {\cal A}_{LR}$ 
at the upgraded Tevatron and the LHC within the MSSM for 
$m_{\tilde t_1}=$160 GeV and 
different values of $\Phi_{\tilde t},
m_{\tilde b_L}$ and $M_{H^{\pm}}$ (with $\mu=120$ GeV and
$M_2=3|\mu|$).} 
\label{fig:four}
\end{figure}

\begin{figure}
\begin{center}
\setlength{\unitlength}{1cm}
\setlength{\fboxsep}{0cm}
\begin{picture}(16.4,18)
\put(-3,-6.5){\epsfbox{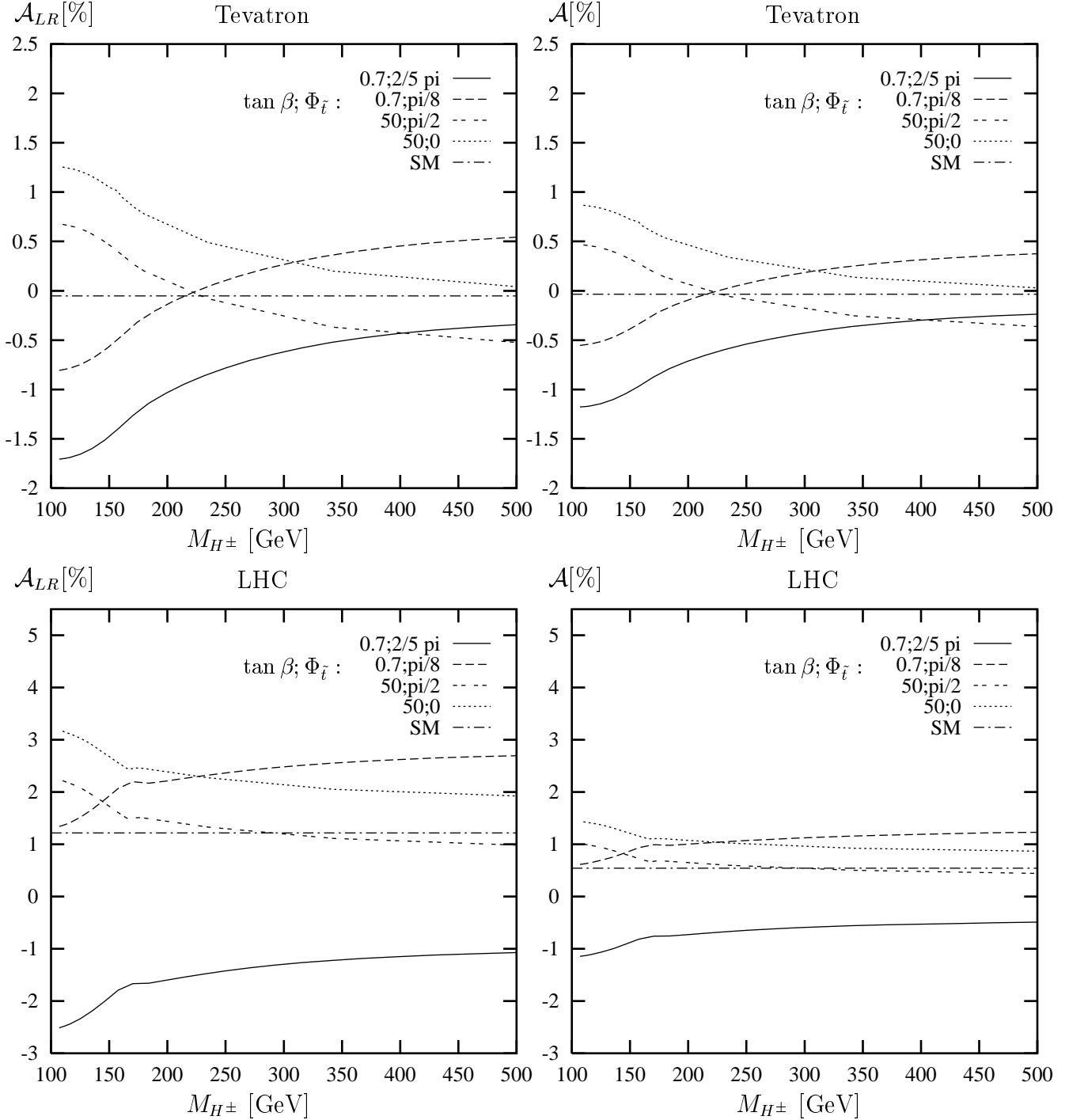}}
\end{picture}
\end{center}
\caption{The variation of the integrated asymmetries ${\cal A}_{LR}$ 
and ${\cal A}$ with $M_{H^{\pm}}$ at the upgraded Tevatron and the LHC
within the MSSM for different values of $\tan\beta$, $\Phi_{\tilde t}$
(with $m_{\tilde t_1}$=90 GeV, $m_{\tilde b_L}$=150 GeV 
($m_{\tilde b_L}$=800 GeV for $\Phi_{\tilde t}=\pi/8$), $\mu=120$ GeV
and $M_2=3|\mu|$).}
\label{fig:five}
\end{figure}

\begin{figure}
\begin{center}
\setlength{\unitlength}{1cm}
\setlength{\fboxsep}{0cm}
\begin{picture}(16.4,18)
\put(-3,-6.5){\epsfbox{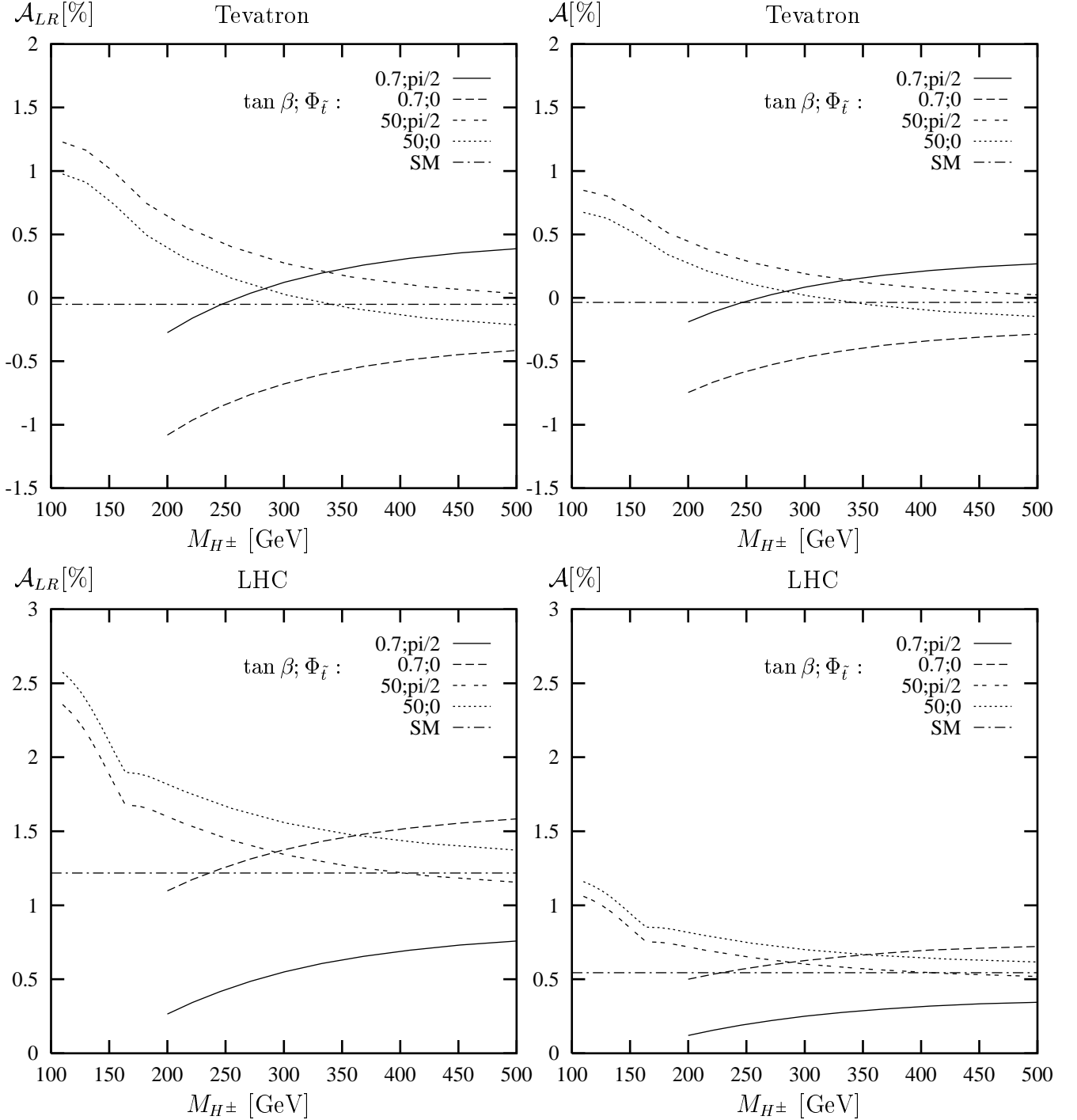}}
\end{picture}
\end{center}
\caption{The variation of the integrated asymmetries ${\cal A}_{LR}$ 
and ${\cal A}$ with $M_{H^{\pm}}$ at the upgraded Tevatron and the LHC
within the MSSM for different values of $\tan\beta$, $\Phi_{\tilde t}$
(with $m_{\tilde t_1}$=160 GeV, $m_{\tilde b_L}$=1150 GeV, $\mu=120$
GeV and $M_2=3|\mu|$).}
\label{fig:six}
\end{figure}

The parity violating asymmetries within the MSSM result from the
interplay of the contribution from the supersymmetric Higgs sector
($H^{\pm}$)and the genuine SUSY EW contribution ($\tilde
\chi^{\pm}_i,\tilde \chi^0_i$).  Depending on the choice of the MSSM
parameters the latter can either enhance or diminish the parity
violating effects loop-induced by the charged Higgs boson
contribution.  This is illustrated with Fig.~\ref{fig:three}
%
%
and Fig.~\ref{fig:four},
where we display the dependence of $\delta {\cal A}_{LR}$ on
$\Phi_{\tilde t}$ and on the mass of the light top squark $m_{\tilde
t_1}$ for the two extreme choices of $\tan\beta$, $\tan\beta=0.7$ and
$\tan\beta=50$. As can be seen from the comparison with the
differential asymmetries obtained within the 2HDM (see
Fig.~\ref{fig:one}) the parity violating SUSY EW contribution is small
for $\Phi_{\tilde t}=\pi/4$, but significant for $\Phi_{\tilde t}=0,
\pi/2$.  This can be easily understood when studying the structure of
the couplings $g_{s,p}^j$ (see Table II of Ref.~\cite{topmssm})
involved in the chargino/neutralino contribution of
Eqs.~(\ref{eq:bten},\ref{eq:bfourteen}), where the dominant Higgsino
part is proportional to $\cos 2\Phi_{\tilde t}$.  The choices for
$\tan\beta$ and $m_{\tilde t_1}$ also affect the relative sign of the
charged Higgs boson and chargino/neutralino contribution, so that in
%
%
Fig.~\ref{fig:three} and Fig.~\ref{fig:four} 
we observe either considerable enhancements or cancellations in
$\delta {\cal A}_{LR}$.  The corresponding integrated asymmetries are
presented in Fig.~\ref{fig:five} and Fig.~\ref{fig:six}, where we show
the variation of ${\cal A}_{LR}$ and ${\cal A}$ with $M_{H^{\pm}}$ for
different MSSM scenarios.
The range for allowed values of $M_{H^{\pm}}$ is constrained by the
LEP 2 experimental limit on the mass of the lighter CP-even Higgs
boson ($M_h$) if $\tan\beta$ is small.
For $\tan\beta=0.7$ and light top squarks we found that the radiative
corrections to the supersymmetric Higgs mass relations always render
$M_h<70$ GeV when the trilinear coupling $A_t = 0$ ($\Phi_{\tilde t} =
0$ or $\pi/2$) in the top squark sector.  Thus, we choose
$\Phi_{\tilde t}=\pi/8, 2\pi/5$ for the light stop scenario
($m_{\tilde t_1}$=90 GeV).  For heavier top squarks and
$\tan\beta=0.7$, $M_h$ can become smaller than $70$ GeV with $M_A \alt
180$ GeV.  Therefore the corresponding curves start only at
$M_{H^{\pm}} \sim 200$ GeV in Fig.~\ref{fig:six}.
At the upgraded Tevatron the parity violating effects within the MSSM
can result in integrated left-right asymmetries of $|{\cal A}_{LR}|\le
1.7\%$ and $|{\cal A}| \le 1.2\%$, which still might be very difficult
to observe.  At the LHC the integrated asymmetries can be as large as
$|{\cal A}_{LR}| \le 3.2\%$ and $|{\cal A}| \le 1.4\%$.

As a sensible measure for the observability of the integrated
left-right asymmetries we follow Refs.~\cite{ttsm,chung} and introduce
a statistical significance $N_S$
\begin{equation}\label{nstat}
N_S=\frac{|{\cal A}_{(LR)}|}{\Delta{\cal A}_{(LR)}} \; .
\end{equation}
In the derivation of $N_S$ we define the observable integrated
left-right asymmetries and their statistical uncertainties
$\Delta{\cal A}_{(LR)}$ as follows
\begin{equation}
{\cal A}_{LR} = \frac{N_{RL}-N_{LR}}{N_{RL}+N_{LR}}
\;  \mbox{ and } \;
{\cal A}  = \frac{N_{R}-N_{L}}{N_{R}+N_{L}} 
\end{equation}
\begin{equation}
\Delta{\cal A}_{LR} =  2 \sqrt{N_{RL} N_{LR}} \,  
(N_{RL}+N_{LR})^{-3/2} \;  \mbox{ and } \;
\Delta{\cal A} =  2 \sqrt{N_{R} N_{L}} \,  
(N_{R}+N_{L})^{-3/2} 
\end{equation}
and approximate the number of the observed polarized $t \bar t$ events
by using
\begin{equation}
N_{LR,RL}  \approx  {\cal L} \, \sigma_{\mp \frac{1}{2},\pm \frac{1}{2}}
\; \mbox{ and } \; 
N_{L,R} \approx  {\cal L} \, (\sigma_{\mp \frac{1}{2},\pm \frac{1}{2}}
+\sigma_{\mp \frac{1}{2},\mp \frac{1}{2}}) \; ,
\end{equation} 
where ${\cal L}$ denotes the integrated luminosity.  

At the LHC, statistics is not the limiting factor:~with ${\cal L}=10
\mbox{ fb}^{-1}$ left-right asymmetries as small as $0.1\%$ are
already statistical significant.  At the Tevatron, however, it might
be difficult to observe even the enhanced effects of the MSSM parity
violating EW interactions.  In Table I we provide the integrated
left-right asymmetries ${\cal A}_{LR}, {\cal A}$ within the 2HDM and
MSSM for different choices of the input parameters together with the
corresponding $N_S$ at the upgraded Tevatron 
%
%
with ${\cal L}=30 \mbox{ fb}^{-1}$ \cite{topacc}.
%
%

Since QCD conserves parity, we have not included SM QCD radiative
corrections in our analysis.  SM QCD corrections might modify the
normalization of the polarization asymmetries, however, they will not
change our conclusions on the observability of parity violating
effects in $t \bar t$ production.  While QCD corrections to strong $t
\bar t$ production might slightly reduce the asymmetries, the number
of observed polarized $t \bar t$ events is expected to be
enhanced. The QCD effect of a slight reduction in the asymmetries can
be partially compensated in the statistical significance $N_S$ of
Eq.~(\ref{nstat}).
%
%

Experimentally, the spin information of the top quark can be obtained
from the angular distribution of the leptons in the top quark decay
\cite{stephen}.  Assuming that the acceptance cuts reduce the cross
section by a factor of 2 and folding in the branching ratio of $B(t
\to W^+ b\to l^+ \nu b \sim 0.214)$ \cite{ewpara}, we obtain a
statistical significance of the integrated asymmetry about $1/3$ of
the value presented in this paper.  However, it has been pointed out
that the parity asymmetry could be significantly enhanced with a cut
on $M_{t\bar{t}}$ \cite{chung}.  With optimal cuts on the
$M_{t\bar{t}}$ and on the angle between the lepton and the top quark,
the parity asymmetries might be visible at the LHC.  A detailed study
at the lepton level with gluon radiation is under investigation and
will be reported in the future.

\section{Conclusions}

We studied the (virtual) effects of parity violating EW interactions
within the 2HDM and the MSSM in polarized $t \bar t$ production at
future hadron colliders:~the upgraded Tevatron $p \bar p$ collider
with $\sqrt{S}=2$ TeV and the LHC $p p$ collider with $\sqrt{S}=14$
TeV.  We calculated the resulting asymmetries in the total production
rate and the invariant mass distribution of left- and right-handed top
quark pairs, ${\cal A},{\cal A}_{LR}$ and $\delta {\cal A}_{LR}$, and
discussed their numerical significance in dependence of the input
parameters of the models under consideration.  While the SM parity
violating EW ${\cal O}(\alpha)$ corrections induce only very small
polarization asymmetries, there could be significant enhancements
within the 2HDM and the MSSM.  In particular, these loop-induced
asymmetries are sensitive to the charged Higgs boson and/or the
top-stop(sbottom)-neutralino(chargino) interaction due to enhanced
Yukawa top-Higgs (Higgsino) couplings.  We find that at the upgraded
Tevatron with ${\cal L}=30 \mbox{ fb}^{-1}$ even non-standard parity
violating effects are still difficult to be observed.  In contrast at
the LHC, statistics is not the limiting factor and polarization
asymmetries as small as $0.1 \%$ are expected to be observable.
Within the 2HDM the loop-induced integrated asymmetries ${\cal
A}_{LR},{\cal A}$ are most pronounced for a light charged Higgs boson
and very small or very large values of $\tan\beta$ and can reach up to
$|{\cal A}_{LR}| \approx 1.4\%, |{\cal A}| \approx 0.9\% $ at the
upgraded Tevatron and up to $|{\cal A}_{LR}| \approx 2.5\%, |{\cal A}|
\approx 1.1\% $ at the LHC.  Within the MSSM the contribution from the
charged Higgs boson can be enhanced or diminished depending on the
mass of the light stop $m_{\tilde t_1}$ and the value of the stop
mixing angle $\Phi_{\tilde t}$. The largest effects can be observed if
there exists a light top squark. 
For instance, for $m_{\tilde t_1}=90$ GeV and $\tan\beta=0.7$ 
the MSSM EW one-loop corrections can induce
left-right asymmetries of $|{\cal A}_{LR}| \le 1.7\%, |{\cal A}|\le
1.2\%$ ($\Phi_{\tilde t}=2\pi/5 $) at the upgraded Tevatron and of
$|{\cal A}_{LR}|\le 2.7\%, |{\cal A}| \le 1.2 \%$ ($\Phi_{\tilde
t}=\pi/8$) at the LHC.  For large values of $\tan\beta$,
$\tan\beta=50$ ($\Phi_{\tilde t}=0$), the integrated left-right
asymmetries can still amount to $|{\cal A}_{LR}| \le 1.3\%, |{\cal
A}|\le 0.9\%$ at the upgraded Tevatron and to $|{\cal A}_{LR}| \le
3.2\%, |{\cal A}|\le 1.4\%$ at the LHC.

\acknowledgements

We are grateful to Howie Haber for beneficial discussions 
and comments on the manuscript.
This research was supported in part by the U.S. Department of Energy
under Grants No. DE-FG02-95ER40896 and in part by the University of
Wisconsin Research Committee with funds granted by the Wisconsin
Alumni Research Foundation.

\newpage
\begin{appendix}

\section{The Born spin amplitudes}
\setcounter{equation}{0}\setcounter{footnote}{0}

We define the spin four-vectors $s_{t,\bar t}$ in the particle's rest
frame in terms of a spin angle $\xi$ as it is illustrated in
Ref.~\cite{yael}
\begin{equation}\label{eq:aone}
s'_{t,\bar t} = (0; \sin\xi,0,\cos\xi) \; ,
\end{equation}
so that in the parton CMS the spin four-vectors read
\begin{eqnarray}\label{eq:atwo}
s_t &=& 
\frac{1}{\gamma_t} (-\beta_t \cos\xi; -\gamma_t \sin\xi,0,-\cos\xi)
\nonumber \\
s_{\bar t} &=& 
\frac{1}{\gamma_t} (-\beta_t \cos\xi; \gamma_t \sin\xi,0,\cos\xi)
\end{eqnarray}
with $\gamma_t=\sqrt{1-\beta_t^2}$.  After applying the projection
operators of Eq.~(\ref{eq:one}) the spin amplitudes are given in terms
of the scalar products of the spin four-vectors and the four-momenta
$p_i$.  In the parton CMS with the $z$-axis chosen to be along the top
quark direction of motion the four-momenta read as follows:
\begin{eqnarray}\label{eq:athree}
p_2 & = & \frac{\sqrt{\hat s}}{2} (1; 0,0,\beta_t) 
\nonumber\\
p_1 & = & \frac{\sqrt{\hat s}}{2} (1; 0,0,-\beta_t) 
\nonumber\\
p_4 & = & \frac{\sqrt{\hat s}}{2} 
(1; \sin\hat\theta,0, \cos\hat \theta) 
\nonumber\\
p_3 & = & \frac{\sqrt{\hat s}}{2} 
(1; -\sin\hat\theta,0,-\cos\hat \theta) \; .
\end{eqnarray}
At the parton level, according to Eq.~(\ref{eq:two}) the polarized
Born differential cross sections $d \sigma_B^i/ d\cos\hat\theta$ for a
generic spin basis are then determined by the following spin
amplitudes (with $z=\cos\hat \theta$)
\begin{eqnarray}\label{eq:afour}
\overline{\sum} \mid {\cal M}_B^{q\bar q}\mid^2
(\lambda_t,\lambda_{\bar t}) 
& = & (4 \pi \alpha_s)^2 \frac{2}{9} \frac{1}{4}
\left\{2-\beta_t^2+\beta_t^2 z^2 \right.
\nonumber\\
&+& \left. 4\lambda_t\lambda_{\bar t} \left[-1+(1-\beta_t^2-2 z^2
+\beta_t^2 z^2)\cos2\xi
-2 z \sqrt{1-z^2} \gamma_t \sin2\xi\right] \right\}
\\ \label{eq:afive}
\overline{\sum} \mid {\cal M}_B^{gg} \mid^2 
(\lambda_t,\lambda_{\bar t}) & = & 
(4\pi \alpha_s)^2 \frac{1}{64} \frac{1}{3} {\cal Y}(\beta_t,z) 
\left\{1+2 \beta_t^2-2\beta_t^4-2\beta_t^2 z^2
+2\beta_t^4 z^2-\beta_t^4 z^4 \right.
\nonumber\\
&+& \left. 4 \lambda_t \lambda_{\bar t} 
\left[1-2 \beta_t^2+\beta_t^2 z^2-
\beta_t^2 \left(-2+3 z^2-2z^4 \right.
\right.\right. 
\nonumber\\
&+& \left. \left. \left. \beta_t^2 (2-2 z^2+z^4)\right) \cos 2\xi
-2 z \beta_t^2 \gamma_t (1-z^2)^{3/2} 
\sin 2\xi \right]\right\} \; ,
\end{eqnarray}
where ${\cal Y}$ is a common spin-independent angular factor as it is
defined in Ref.~\cite{stephen}
\begin{equation}
{\cal Y}(\beta_t,z) = \frac{7+9\beta_t^2 z^2}{(1-\beta_t^2z^2)^2} \; .
\end{equation}
For $\xi=\pm \pi$ the helicity basis is recovered where the particle's
spin is decomposed along its direction of motion.  As discussed in
Refs.~\cite{stephen,yael} the freedom to choose $\xi$ can be used to
enhance spin correlations in the production of polarized $t\bar t$
pairs.

\section{The form factors}
\setcounter{equation}{0}\setcounter{footnote}{0}

After performing an {\em on-shell} renormalization procedure as
described in Ref.~\cite{diplpubsm} the ${\cal O}(\alpha)$ contribution
to strong top pair production is parameterized in terms of UV finite
form factors.  Depending on the choice of the model (SM, 2HDM or MSSM)
the renormalized vector form factor ($F_V$), the magnetic form factor
($F_M$) and the axial vector form factor ($G_A$) comprise the weak
gauge boson contribution $F_{V,M}^G,G_A^G$, the Higgs boson
contribution $F_{V,M}^H,G_A^H$ and/or the genuine SUSY contribution
$S_{V,M,A}$
\begin{eqnarray}\label{eq:bone}
F_{V,M}(\hat s,m_q) &=& \left(\frac{m_q}{2 s_w M_W}\right)^2 \, 
\sum_{S=\eta,\chi,\Phi^{\pm}\atop S=H^0,h^0,A^0,H^{\pm},G^0,G^{\pm}} 
F_{V,M}^H(\hat s,m_q,M_S)+\sum_{V=Z^0,W^{\pm}} 
F_{V,M}^G(\hat s,m_q,M_V) \; .
\nonumber\\
&+& \sum_{j=1}^2 \left(\sum_{i=1}^4 
S_{V,M}(m_{\tilde q_j},M_{\tilde \chi_i^0})
+ \sum_{i=1}^2 S_{V,M}(m_{\tilde {q}'_j},M_{\tilde \chi_i^{\pm}})\right)
\end{eqnarray}
\begin{eqnarray}\label{eq:btwo}
G_A(\hat s,m_q) & = & \left(\frac{m_q}{2 s_w M_W}\right)^2 \, 
\sum_{S=\Phi^{\pm}\atop S=H^{\pm},G^{\pm}} G_A^H(\hat s,m_q,M_S)
+\sum_{V=Z^0,W^{\pm}} G_A^G(\hat s,m_q,M_V) 
\nonumber\\
&+& \sum_{j=1}^2 \left(\sum_{i=1}^4 
S_A(m_{\tilde q_j},M_{\tilde \chi_i^0})
+ \sum_{i=1}^2 S_A(m_{\tilde {q}'_j},M_{\tilde \chi_i^{\pm}})\right) \; ,
\end{eqnarray}
where $\tilde{q}'_j$ denotes the isospin partner of $\tilde{q}_j$.  In
case of final state vertex corrections $m_q=m_t$ and $m_{\tilde
q_j}=m_{\tilde t_j}$.  For the initial state contribution to the
$q\bar q$ annihilation subprocess we assume $m_q=0$, so that there is
no Higgs boson contribution and $(F,S)_M=0$. For the genuine initial
state SUSY contribution the masses of the supersymmetric partners of
the light quarks $m_{\tilde q_j}$ are fixed by choosing $m_{\tilde
b_j}$ as it is described in Ref.~\cite{topmssm}.  Explicit expressions
for $F_{V,M}$ and $S_{V,M}$ are provided in
Refs.~\cite{diplpubsm,topmssm}.  The renormalized axial vector form
factors read
\begin{eqnarray}\label{eq:bthree}
G_A^H(\hat s,m_t,M_S) & = & c_s (c_p+{c'}_p) \,
[ -(m'^2-m_t^2) \, C_0 + 2 C_2^0 -\frac{1}{2}-4 m_t^2 \, C_1^+
\nonumber\\
&+& \hat s \, C_2^- +(4 m_t^2-\hat s)\, C_2^+](\hat s,m_t,m_t,M_S)
-\delta Z_A^H 
\nonumber \\
G_A^G(\hat s,m_q,M_V) & = & 4 g_V g_A \, [(\hat s+m'^2-m_q^2)\, C_0+1
-2 C_2^0-2 (\hat s-2 m_q^2)\, C_1^+
\nonumber\\
&-& \hat s \, C_2^- -(4 m_t^2-\hat s)\, C_2^+](\hat s,m_q,m_q,M_V)
-\delta Z_A^G
\nonumber \\
S_A(m_1,m_2)& = & 2 g_s^j g_p^j \, 2 C_2^0(\hat s,m_1,m_1,m_2)
-\delta Z_A^S \; .
\end{eqnarray}
In the diagrams involving the charged Higgs boson and the $W$ boson,
$m'$ denotes the mass of the isospin partner of $q,t$, and $m' =
m_{q,t}$ in the contributions of neutral Higgs bosons and the $Z$
boson.  The top Yukawa couplings $c_s,c_p,c'_p$ are described in
Refs.~\cite{diplpubsm} (SM) and~\cite{topmssm} (2HDM). The quark-gauge
boson couplings $g_V,g_A$ are given in Ref.~\cite{diplpubsm}.
$g_s^j,g_p^j$ denote the scalar and pseudo-scalar coupling,
respectively, describing the
neutralino(chargino)-stop(sbottom)-top-vertex (Table II of
Ref.~\cite{topmssm}).  The renormalization constant $\delta
Z_A^{G,H,S}$ is defined by the axial vector part of the unrenormalized
quark self energy
\begin{equation}\label{eq:bfour}
\delta Z_A^{G,H,S} = -\Sigma_A^{G,H,S}(p^2 = m_q^2) \; .
\end{equation}
The weak gauge boson and Higgs boson contribution $\Sigma_A^{G,H}$
is explicitly given in Ref.~\cite{diplpubsm}. 
The EW  SUSY contribution reads 
\begin{equation}\label{eq:bfive}
\Sigma_A^S(p^2) = 2 g_s^j g_p^j \, B_1(p^2,m_2,m_1) \; .
\end{equation}
We follow the notations of the $B,C$-functions in
Ref.~\cite{diplpubsm}.

The ${\cal O}(\alpha)$ contribution to the $t(u)$-channel of the gluon
fusion subprocess comprises the parity conserving part parameterized
in terms of the form factors $\rho_i^{X,(t,u)}$
\cite{diplpubsm,topmssm}
\begin{eqnarray}\label{eq:bsix}
\rho_i^{(V,\Sigma,\Box),(t,u)} & = & 
\left(\frac{m_t}{2 s_w M_W}\right)^2 \,
\sum_{S} H_i^{(V,\Sigma,\Box),(t,u)}
+\sum_V G_i^{(V,\Sigma,\Box),(t,u)} 
\nonumber\\
& + & \sum_{j=1}^2 \left(\sum_{k=1}^4 T_i^{(V,\Sigma,\Box),(t,u)}
(m_{\tilde t_j},M_{\tilde \chi_k}) 
+ \sum_{k=1}^2 T_i^{(V,\Sigma,\Box),(t,u)}
(m_{\tilde b_j},M_{\tilde \chi_k})\right)
\nonumber \\
\rho_{12}^{\triangleleft} & = & \left(\frac{m_t}{2 s_w M_W}\right)^2
 \, H_{12}^{\triangleleft} \; ,
\end{eqnarray}
and the parity violating part
\begin{eqnarray}\label{eq:bseven}
\sigma_i^{(V,\Sigma,\Box),(t,u)} 
& = & \left(\frac{m_t}{2 s_w M_W}\right)^2 \,
\sum_{S} {\cal H}_i^{(V,\Sigma,\Box),(t,u)}
+\sum_V {\cal G}_i^{(V,\Sigma,\Box),(t,u)}
\nonumber\\
&+&  \sum_{j=1}^2 \left(\sum_{k=1}^4 
{\cal T}_i^{(V,\Sigma,\Box),(t,u)}
(m_{\tilde t_j},M_{\tilde{\chi}_k^0}) 
+ \sum_{k=1}^2 {\cal T}_i^{(V,\Sigma,\Box),(t,u)}
(m_{\tilde b_j},M_{\tilde{\chi}_k^{\pm}})\right) \; .
\end{eqnarray}
There the parity violating parts of the weak gauge boson, Higgs boson
and genuine SUSY contributions to the vertex corrections, the top
quark self-energy insertions and the box diagrams are described by
$({\cal G},{\cal H},{\cal T})_i^{(V,\Sigma,\Box),(t,u)}$,
respectively.\\
      
\underline{Vertex corrections:}
\begin{eqnarray}\label{eq:beight}
{\cal H}_1^{V,t} &=&  2 c_s (c_p+{c'}_p) \left[B_0(0,m',m')-
2 C_2^0-(m'^2-m_t^2-M_S^2)\, C_0
\right. \nonumber\\
&+& \left. (\hat t+m_t^2)\, C_1^2+2 m_t^2 C_1^1\right] 
-2 \delta Z_A^H
\nonumber\\
{\cal H}_4^{V,t} & = & 2 c_s (c_p+{c'}_p) (m_t^2-\hat t) 
\, [C_1^2+C_2^2+C_2^{12}]
\nonumber\\
{\cal H}_{14}^{V,t} & = & 2 c_s (c_p+{c'}_p) 
\, [-C_1^1+C_1^2-C_2^1+C_2^2]
\end{eqnarray}
and 
\begin{eqnarray}\label{eq:bnine}
{\cal G}_1^{V,t} &=& 8 g_V g_A \left[(m'^2-M_V^2-\hat t)
\, C_0-B_0(0,m',m')+2 C_2^0+\frac{1}{2}
\right. \nonumber\\
&-& \left. (\hat t+m_t^2)\, C_1^1-2 \hat t C_1^2\right] 
-2 \delta Z_A^G
\nonumber\\
{\cal G}_4^{V,t} & = & 8 g_V g_A (\hat t-m_t^2) 
\, [C_0+C_1^1+2 C_1^2+C_2^2+C_2^{12}]
\nonumber\\
{\cal G}_{14}^{V,t} & = & 8 g_V g_A 
\, [C_1^1-C_1^2+C_2^1-C_2^2] \; ,
\end{eqnarray}
where the three-point functions $C(\hat t,m',m',M_{V,S})$ are given in
Ref.~\cite{diplpubsm}.
\begin{eqnarray}\label{eq:bten}
{\cal T}_1^{V,t}(m_1,m_2) & = & 8 g_s^j g_p^j 
\, C_2^0(\hat t,m_1,m_1,m_2)-2 \delta Z_A^S
\nonumber\\
{\cal T}_4^{V,t}(m_1,m_2) & = & 4 g_s^j g_p^j (\hat t-m_t^2)
(C_2^2+C_2^{12}+C_1^2)(\hat t,m_1,m_1,m_2)
\nonumber\\
{\cal T}_{14}^{V,t}(m_1,m_2) & = & -4 g_s^j g_p^j
(-C_2^1+C_2^2-C_1^1+C_1^2)(\hat t,m_1,m_1,m_2)
\end{eqnarray}

\underline{Top quark self-energy insertion:}

\begin{equation}\label{eq:beleven}
({\cal H},{\cal G},{\cal T})^{\Sigma,t}_1(m_1,m_2) 
= (\hat t-m_t^2) \, 
[\Sigma_A^{(H,G,S)}(\hat t,m_1,m_2)+\delta Z_A^{(H,G,S)}] 
\end{equation}

\underline{Box contribution:}

\begin{eqnarray}\label{eq:btwelve}
{\cal H}_1^{\Box,t} &=& c_s (c_p+{c'}_p) 
\, \left[C_0+(M_S^2+m_t^2-m'^2)\, D_0
+(m'^2+m_t^2)\, D_1^2-6\, D_2^0 \right.
\nonumber\\
&-& \left. \hat t \, (2 D_2^2+D_3^2)+4 m_t^2\, D_1^1
-2 (\hat t+m_t^2)\, (D_2^{12}+D_3^{21})
-6\, D_3^{02}-m_t^2\, 2 D_3^{12}
\right.
\nonumber\\
&+& \left. (\hat s-2 m_t^2) \, D_3^{123} \right]
\nonumber\\
{\cal H}_2^{\Box,t} &=& -2 c_s (c_p+{c'}_p) \, [D_2^0+D_3^{02}]
\nonumber\\
{\cal H}_4^{\Box,t} &=& c_s (c_p+{c'}_p) 
\left[2 D_2^0+m_t^2\, (2 (D_2^1+D_2^{13})+D_3^1) \right.
\nonumber\\
&+&\left. (m_t^2-m'^2)\, (D_1^1+D_1^2)+(\hat t+5 m_t^2)\, D_2^{12}
+2 D_3^{01}+4 D_3^{02}+\hat t \, D_3^22+(\hat t+3 m_t^2)\, D_3^{12}
\right.
\nonumber\\
&+& \left. (3 \hat t+2 m_t^2)\, D_3^{21}+(3 m_t^2-\hat s)\, D_3^{13}
+(\hat t-\hat s+3 m_t^2) \, D_3^{123}+\hat t \, D_2^2 \right]
\nonumber\\
{\cal H}_6^{\Box,t} &=& 2 c_s (c_p+{c'}_p) 
\, [D_2^2+2 D_2^{12}+D_3^2+2 D_3^{12}
+4 D_3^{21}+2 D_3^{123}]
\nonumber\\
{\cal H}_{14}^{\Box,t} &=& 2 c_s (c_p+{c'}_p) 
\, [D_1^2+2 D_2^{12}+D_2^2]
\end{eqnarray}
and
\begin{eqnarray}\label{eq:bthirteen}
{\cal G}_1^{\Box,t} & = & 4 g_V g_A \, \left[C_0 +
(m_t^2-m'^2+\hat s+M_V^2) \, D_0-2\,(\hat s+2 m_t^2)\, D_1^1
\right. \nonumber \\
& + & \left. (\hat s+m'^2+m_t^2) \, D_1^2 -6 D_2^0
- \hat t \, (2 D_2^2+ D_3^2)
\right. \nonumber \\
&-& \left. 2\, (\hat t+m_t^2) \,(D_2^{12}+D_3^{21})
-6 D_3^{02}-2 m_t^2 D_3^{12}+(\hat s-2m_t^2)\,D_3^{123}\right]
\nonumber\\
{\cal G}_2^{\Box,t} & = & 4 g_V g_A 
\, \left[C_0+(m_t^2-m'^2+M_V^2) \, D_0
+(m'^2+m_t^2) \, D_1^2 \right.
\nonumber \\
&-& \left. 4 D_2^0- \hat t \, (2 D_2^2+ D_3^2)
-2\, (\hat t+m_t^2) \, (D_2^{12}+D_3^{21})\right.
\nonumber \\
&-& \left. 4 D_3^{02}-2 m_t^2 (D_3^{12}-2 D_1^1)
+(\hat s-2m_t^2)\,D_3^{123}\right]
\nonumber \\
{\cal G}_4^{\Box,t} & = & 8 g_V g_A \, 
\left[-C_0+(M_V^2+m_t^2-m'^2)\, D_0
-(5m_t^2-m'^2-\hat s)\, D_1^1 \right.
\nonumber\\
&-& \left. (\hat t+m_t^2)\,(D_1^2+D_3^{12}+D_3^{123})+4 D_2^0 
\right.
\nonumber \\
& + & \left. (\hat s-\hat t)\,D_2^{12}+(\hat s-2m_t^2)\, 
(D_2^1+ D_2^{13}) \right.
\nonumber \\
& - & \left. 2 (D_3^{01}- D_3^{02}) - m_t^2 \, D_3^1
-\hat t\, D_3^{21}-(3m_t^2-\hat s) \, D_3^{13} \right]
\nonumber \\
{\cal G}_6^{\Box,t} & = & -8 g_V g_A 
\, \left[2 D_1^2+6 D_2^{12}+3 D_2^2
+D_3^2+2 D_3^{12}+4 D_3^{21}+2 D_3^{123}\right]
\nonumber \\
{\cal G}_{14}^{\Box,t}& = & 8 g_V g_A 
\, [D_1^2+D_2^2+2 D_2^{12}] \; ,
\end{eqnarray}
where the three- and four-point functions are denoted by $C_0 =
C_0(\hat s,m',m',m')$ and $[D_0,D_i^j] = [D_0,D_i^j](\hat
t,m',m',m',M_{S,V})$, respectively.

\begin{eqnarray}\label{eq:bfourteen}
{\cal T}_2^{\Box,t}(m_1,m_2) & = & 4 g_s^j g_p^j 
D_3^{02}(\hat t,m_1,m_1,m_1,m_2)
\nonumber\\
{\cal T}_4^{\Box,t}(m_1,m_2) & = & 8 g_s^j g_p^j 
(D_2^0+2 D_3^{01}+D_3^{02})(\hat t,m_1,m_1,m_1,m_2)
\nonumber\\
{\cal T}_6^{\Box,t}(m_1,m_2) & = & -4 g_s^j g_p^j
(D_1^2+D_3^2+2 (2 D_2^{12}+D_2^2+D_3^{12}+D_3^{123}+2 D_3^{21}))
(\hat t,m_1,m_1,m_1,m_2) \; . 
\nonumber\\
\end{eqnarray}
The $u$-channel contribution $\sigma_i^{(V,\Sigma,\Box),u}$ can be
obtained from the $t$-channel form factors by replacing $\hat t$ with
$\hat u$.

\section{The interference with the Born matrix element}
\setcounter{equation}{0}\setcounter{footnote}{0}

The polarized $gg\rightarrow t\bar t$ cross section of
Eq.~\ref{eq:eight} is given in terms of the form factors and the
factors $M^{(V,A),(t,u)}_{\lambda_t\lambda_{\bar t}}(i,j)$
($i=1,\cdots 17,j=1,2,3$) which result from the interference of the
standard matrix elements with the Born matrix elements according to
Eq.~(\ref{eq:nine}).  Here we only provide explicit expression for
those factors which multiply the non-vanishing form factors
$F_{V,M},G_A$ and $(\rho,\sigma)_i^{X,(t,u)}$ of Appendix B.  Using
the gluon polarization sum of Eqs.~(2.12,2.13) in
Ref.~\cite{diplpubsm} these factors read as follows (with $z=\cos\hat
\theta, \gamma_t^2=1-\beta_t^2$ and using the helicity basis):
\begin{eqnarray}\label{eq:cone}
M^{V,t}_{\lambda_t\lambda_{\bar t}}(1,1) & = & 
(1-\beta_t z) [-1+\beta_t^2 z^2 
+ 4 \lambda_t \lambda_{\bar t} (1-2 z^2+\beta_t^2 z^2)] \hat s/ 8
\nonumber\\
M^{V,t}_{\lambda_t\lambda_{\bar t}}(1,2) & = & 
[-2+(1+4\lambda_t\lambda_{\bar t})(\beta_t^4 
+4 \beta_t^2 z^2-3 \beta_t^4  z^2- 2 \beta_t^3  z^3  
\nonumber\\
&+& 2 \beta_t^4 z^4-z^2+2 \beta_t z^3-2 \beta_t^2 z^4) 
\nonumber\\
&+& (1-4 \lambda_t \lambda_{\bar t}) 
(-\beta_t^2 + 2 \beta_t z -\beta_t^2 z^2+z^2-2\beta_t z^3
+2 \beta_t^2 z^4)]\hat s/[8 (1-\beta_t z)]
\nonumber\\
M^{V,t}_{\lambda_t\lambda_{\bar t}}(1,3) & = &  
(1-z^2) [-\beta_t^2 +\beta_t^4 -2 \beta_t^4 z^2
\nonumber\\
&+& 4 \lambda_t \lambda_{\bar t} (-2+\beta_t^2+ \beta_t^4 
+4 \beta_t^2 z^2 -2 \beta_t^4 z^2)]\hat s/[8 (1+\beta_t z)]
\nonumber\\
\\
M^{V,t}_{\lambda_t\lambda_{\bar t}}(2,1) & = & 
[1-\beta_t^2 z^2+4 \lambda_t\lambda_{\bar t} 
(-1+2 z^2-\beta_t^2 z^2)] \hat s/4
\nonumber\\
M^{V,t}_{\lambda_t\lambda_{\bar t}}(2,2) & = & 
[1+(1+4 \lambda_t\lambda_{\bar t}) 
(-\beta_t^3 z-\beta_t^2 z^2+\beta_t^3 z^3)
\nonumber\\
&+& 4 \lambda_t\lambda_{\bar t} (-1+2 \beta_t z+2 z^2-2 \beta_t z^3)]
\hat s/[4 (1-\beta_t z)]
\nonumber\\
M^{V,t}_{\lambda_t\lambda_{\bar t}}(2,3) & = & 
[-1+(1+4 \lambda_t\lambda_{\bar t}) 
(-\beta_t^3 z+\beta_t^2 z^2+\beta_t^3 z^3)
\nonumber\\
&+& 4 \lambda_t\lambda_{\bar t} 
(1+2 \beta_t z-2 z^2-2 \beta_t z^3)]\hat s/[4 (1+\beta_t z)]
\nonumber\\
\\
M^{V,t}_{\lambda_t\lambda_{\bar t}}(4,1) & = & 
(1-z^2) [\beta_t^2+4 \lambda_t \lambda_{\bar t} (-2+\beta_t^2)] 
\beta_t z \hat s/8
\nonumber\\
M^{V,t}_{\lambda_t\lambda_{\bar t}}(4,2) & = & 
(1-z^2) [-2 \beta_t-(1+4\lambda_t \lambda_{\bar t}) 
(\beta_t-2 \beta_t^3-\beta_t^2 z
+2 \beta_t^3 z^2)
\nonumber\\
&-&8 \lambda_t \lambda_{\bar t} (z-2 \beta_t z^2)]\beta_t 
\hat s/[8 (1-\beta_t z)]
\nonumber\\
M^{V,t}_{\lambda_t\lambda_{\bar t}}(4,3) & = & (1-z^2) 
[-2 \beta_t-(1+4\lambda_t \lambda_{\bar t}) 
(\beta_t-2 \beta_t^3+\beta_t^2 z
+2 \beta_t^3 z^2)
\nonumber\\
&-& 8 \lambda_t \lambda_{\bar t} (-z-2 \beta_t z^2)]\beta_t 
\hat s/[8 (1+\beta_t z)]
\nonumber\\
\\
M^{V,t}_{\lambda_t\lambda_{\bar t}}(6,1) & = & (1-z^2) 
[1-\beta_t^2 z^2 -4 \lambda_t \lambda_{\bar t} (1-2 z^2
+\beta_t^2 z^2)] \beta_t^2 \hat s^2/32
\nonumber\\
M^{V,t}_{\lambda_t \lambda_{\bar t}}(6,2) & = & 
(1-z^2) [2-(1+4 \lambda_t \lambda_{\bar t}) (1+2 \beta_t^3 z
+\beta_t^2 z^2-2 \beta_t^3 z^3)
\nonumber\\
&-& 8 \lambda_t \lambda_{\bar t} (-2 \beta_t z-z^2+2 \beta_t z^3)]
\beta_t^2 \hat s^2/[32(1-\beta_t z)]
\nonumber\\
M^{V,t}_{\lambda_t\lambda_{\bar t}}(6,3) & = & 
(1-z^2) [-2-(1+4 \lambda_t \lambda_{\bar t}) (-1+2 \beta_t^3 z
-\beta_t^2 z^2-2 \beta_t^3 z^3)
\nonumber\\
&-&8 \lambda_t \lambda_{\bar t} (-2 \beta_t z+ z^2+2 \beta_t z^3)]
\beta_t^2 \hat s^2/[32(1+\beta_t z)] 
\nonumber\\
\\
M^{V,t}_{\lambda_t\lambda_{\bar t}}(11,1) & = & 
\gamma_t^2 (1+4 \lambda_t\lambda_{\bar t}) \beta_t z \hat s/8
\nonumber\\
M^{V,t}_{\lambda_t\lambda_{\bar t}}(11,2) & = & 
\gamma_t^2 (\beta_t+2 z-\beta_t z^2) (1+4 \lambda_t\lambda_{\bar t})
\beta_t \hat s/[8(1-\beta_t z)]
\nonumber\\
M^{V,t}_{\lambda_t\lambda_{\bar t}}(11,3) & = & 
\gamma_t^2 (1-z^2) (1+4 \lambda_t\lambda_{\bar t})
\beta_t^2 \hat s/[8(1+\beta_t z)] 
\nonumber\\
\\
M^{V,t}_{\lambda_t\lambda_{\bar t}}(12,1) & = & 
\gamma_t^2 (1+4 \lambda_t\lambda_{\bar t})\beta_t z \hat s/8
\nonumber\\
M^{V,t}_{\lambda_t\lambda_{\bar t}}(12,2) & = & \gamma_t^2 
(\beta_t+z-\beta_t z^2) (1+4 \lambda_t\lambda_{\bar t})
\beta_t \hat s/[8(1-\beta_t z)]
\nonumber\\
M^{V,t}_{\lambda_t\lambda_{\bar t}}(12,3) & = & \gamma_t^2 
(\beta_t-z-\beta_t z^2) (1+4 \lambda_t\lambda_{\bar t})
\beta_t \hat s/[8(1+\beta_t z)]
\nonumber\\
\\
M^{V,t}_{\lambda_t\lambda_{\bar t}}(14,1) & = & 
\gamma_t^2  (1-z^2) \lambda_t \lambda_{\bar t} \beta_t z \hat s^2/4
\nonumber\\
M^{V,t}_{\lambda_t\lambda_{\bar t}}(14,2) & = &  
\gamma_t^2 (1-z^2) [\beta_t+4 \lambda_t\lambda_{\bar t} 
(z-2 \beta_t z^2)] \beta_t \hat s^2/[16(1-\beta_t z)]
\nonumber\\
M^{V,t}_{\lambda_t\lambda_{\bar t}}(14,3) & = & 
\gamma_t^2  (1-z^2) [\beta_t+4 \lambda_t\lambda_{\bar t} 
(-z-2 \beta_t z^2)] \beta_t \hat s^2/[16(1+\beta_t z)] 
\nonumber\\
\\
M^{V,t}_{\lambda_t\lambda_{\bar t}}(16,1) & = &  
\gamma_t^2 (1-z^2) (1+4 \lambda_t\lambda_{\bar t}) 
\beta_t^3 z \hat s^2/64
\nonumber\\
M^{V,t}_{\lambda_t\lambda_{\bar t}}(16,2) & = & 
\gamma_t^2 (1-z^2) (2\beta_t+z-2 \beta_t z^2)
(1+4 \lambda_t\lambda_{\bar t})
\beta_t^3 \hat s^2/[64(1-\beta_t z)]
\nonumber\\
M^{V,t}_{\lambda_t\lambda_{\bar t}}(16,3) & = & 
\gamma_t^2 (1-z^2) (2\beta_t-z-2 \beta_t z^2)
(1+4 \lambda_t\lambda_{\bar t})
\beta_t^3 \hat s^3/[64(1+\beta_t z)] \; .
\end{eqnarray}
The parity violating part of the EW one-loop corrections to the gluon
fusion subprocess is determined by the form factors $G_A$ and
$\sigma_i^{X,(t,u)}$ and the factors
$M^{A,(t,u)}_{\lambda_t\lambda_{\bar t}}(i,j)$.  According to
Eq.~(\ref{eq:nine}) these factors yield as follows (with
$\Delta\lambda=\lambda_t-\lambda_{\bar t}$ and using the helicity
basis)
\begin{eqnarray}\label{eq:ctwo}
M^{A,t}_{\lambda_t\lambda_{\bar t}}(1,1)& =& \Delta\lambda (1-z^2) 
(-1+\beta_t z) \beta_t \hat s/4
\nonumber\\
M^{A,t}_{\lambda_t\lambda_{\bar t}}(1,2)& =& \Delta\lambda  (1-z^2) 
(-1-\beta_t^2+2 \beta_t z-2\beta_t^2 z^2) 
\beta_t \hat s/[4(1-\beta_t z)]
\nonumber\\
\\
M^{A,t}_{\lambda_t\lambda_{\bar t}}(2,1)& =& \Delta\lambda (1-z^2) 
\beta_t \hat s/2
\nonumber\\
M^{A,t}_{\lambda_t\lambda_{\bar t}}(2,2)& =&  \Delta\lambda (1-z^2) 
(1-\beta_t z) \beta_t \hat s/[2(1-\beta_t z)]
\nonumber\\
\\
M^{A,t}_{\lambda_t\lambda_{\bar t}}(4,1)& =&  \Delta\lambda (1-z^2) 
z \beta_t^2 \hat s/4
\nonumber\\
M^{A,t}_{\lambda_t\lambda_{\bar t}}(4,2)& =&  \Delta\lambda (1-z^2) 
(-\beta_t+z-2\beta_t z^2) 
\beta_t^2 \hat s/[4(1-\beta_t z)]
\nonumber\\
\\
M^{A,t}_{\lambda_t\lambda_{\bar t}}(6,1)& = & \Delta\lambda 
(1-z^2)^2 \beta_t^3 \hat s^2/16
\nonumber\\
M^{A,t}_{\lambda_t\lambda_{\bar t}}(6,2)& =&  \Delta\lambda 
(1-z^2)^2 (1-2\beta_t z) 
\beta_t^3 \hat s^2/[16(1-\beta_t z)]
\nonumber\\
\\
M^{A,t}_{\lambda_t\lambda_{\bar t}}(14,1)& =&  \Delta\lambda 
(1-z^2) (-1+\beta_t^2) \beta_t \hat s^2/16
\nonumber\\
M^{A,t}_{\lambda_t\lambda_{\bar t}}(14,2)& =& \Delta\lambda 
(1-z^2) (-1+\beta_t^2) (1-3 \beta_t z) 
\beta_t \hat s^2/[16(1-\beta_t z)]
\nonumber\\
\\
M^{A,t}_{\lambda_t\lambda_{\bar t}}(i,3)& =& 
M^{A,t}_{\lambda_t\lambda_{\bar t}}(i,2) (z \to -z) \; .
\end{eqnarray}
The interference of the $u$-channel next-to-leading contribution to
the gluon fusion subprocess with the Born matrix element leads to the
factors $M^{(V,A),u}_{\lambda_t\lambda_{\bar t}}(i,j)$ according to
Eq.~(\ref{eq:nine}). They can be derived from
$M^{(V,A),t}_{\lambda_t\lambda_{\bar t}}(i,j)$ by performing the
following replacements
\begin{eqnarray}\label{eq:ctree}
M^{(V,A),u}_{\lambda_t\lambda_{\bar t}}(i,1) & = & 
-M^{(V,A),t}_{\lambda_t\lambda_{\bar t}}(i,1)(z \rightarrow -z)
\nonumber\\
M^{(V,A),u}_{\lambda_t\lambda_{\bar t}}(i,2) & = & 
M^{(V,A),t}_{\lambda_t\lambda_{\bar t}}(i,3)(z \rightarrow -z)
\nonumber\\
M^{(V,A),u}_{\lambda_t\lambda_{\bar t}}(i,3) & = & 
M^{(V,A),t}_{\lambda_t\lambda_{\bar t}}(i,2)(z \rightarrow -z) \; .
\end{eqnarray}  
The factors to unpolarized $t \bar t$ production $M^{(t,u)}(i,j)$ of
Appendix B.2 in Ref.~\cite{diplpubsm} are recovered when summing the
corresponding polarized factors over the $t$ and $\bar t$ polarization
states $M^{(t,u)}(i,j)=\sum_{\lambda_t,\lambda_{\bar t} }
M^{V,(t,u)}_{\lambda_t\lambda_{\bar t}}(i,j)$. Naturally, the sum of
the parity violating factors $M^{A,(t,u)}_{\lambda_t\lambda_{\bar
t}}(i,j)$ over all the $t$ and $\bar t$ polarizations is zero, so that
there are no effects of loop-induced parity violating interactions in
unpolarized $t\bar t$ production.
\end{appendix}


\newpage

%
%
%
%

\newpage

\renewcommand{\arraystretch}{1.2}
\renewcommand{\baselinestretch}{0.98}
\begin{table}
\caption{The polarization asymmetries ${\cal A}_{LR}, {\cal A}$ 
and their statistical significance $N_S$ within the 2HDM and the MSSM
at the upgraded Tevatron with ${\cal L}=30 \mbox{ fb}^{-1}$.}
\begin{tabular}{|cccc||cccc|}
\multicolumn{8}{c}{2HDM with $\alpha=\pi/2$, $M_H=75$ GeV, 
$M_h=70$ GeV and $M_A=75$ GeV} \\ \hline
$\tan\beta$ & $M_{H^{\pm}}$ [GeV] &  &  &
${\cal A}_{LR} [\%]$ & $N_S$ & ${\cal A} [\%]$ & $N_S$ \\ \hline
0.7 & 100 & & & -1.35 & 4.71 & -0.93  & 3.90  \\
0.7 & 180 & & & -0.83 & 2.88 & -0.57  & 2.39  \\ 
0.7 & 300 & & & -0.29 & 1.00 & -0.20  & 0.83  \\ \hline
2   & 100 & & & -0.21 & 0.73 & -0.14  & 0.61  \\
2   & 180 & & & -0.15 & 0.52 & -0.10  & 0.42  \\ 
2   & 300 & & & -0.08 & 0.28 & -0.06  & 0.23  \\ \hline
50  & 100 & & &  1.12 & 3.88 & 0.77   & 3.22  \\ 
50  & 180 & & &  0.65 & 2.25 & 0.45   & 1.85  \\ 
50  & 300 & & &  0.16 & 0.55 & 0.11   & 0.47  \\ \hline \hline
\multicolumn{8}{c}{MSSM with $\mu=120$ GeV and $M_2=3|\mu|$}\\ \hline
$\tan\beta$ & $M_{H^{\pm}}$ [GeV] & $\Phi_{\tilde t}$ & $m_{\tilde b_L}$ 
[GeV] & ${\cal A}_{LR} [\%]$ & $N_S$ & ${\cal A} [\%]$ 
& $N_S$ \\ \hline
\multicolumn{8}{|c|}{$m_{\tilde t_1}=160$ GeV} \\ \hline
0.7 & 200 & $\pi/2$ & 1150 & -0.27  & 0.97  & -0.19  & 0.80 \\
0.7 & 110 & $\pi/4$ & 350  & -1.18  & 4.10  & -0.81  & 3.41 \\
0.7 & 200 & 0       & 1150 & -1.08  & 3.79  & -0.75  & 3.15 \\ \hline
2   & 160 & $\pi/2$ & 900  & -0.01  & 0.05  & -0.01  & 0.03 \\
2   & 110 & $\pi/4$ & 500  & -0.19  & 0.66  & -0.13  & 0.54 \\
2   & 160 & 0       & 900  & -0.32  & 1.11  & -0.22  & 0.94 \\ \hline
50  & 110 & $\pi/2$ & 800  &  1.22  & 4.23  &  0.84  & 3.50 \\
50  & 110 & $\pi/4$ & 300  &  1.01  & 3.50  &  0.70  & 2.91 \\
50  & 110 & 0       & 800  &  0.97  & 3.38  &  0.67  & 2.81 \\ \hline
\multicolumn{8}{|c|}{$m_{\tilde t_1}=90$ GeV}\\ \hline
0.7 & 110 & $2\pi/5$  & 150  &  -1.70  & 5.86 &  -1.17 & 4.87 \\
0.7 & 110 & $\pi/4$   & 300  &  -1.13  & 3.88 &  -0.78 & 3.22 \\
0.7 & 110 & $\pi/8$   & 800  &  -0.80  & 2.75 &  -0.55 & 2.28 \\ \hline
2   & 210 & $\pi/2$   & 1150 &  -0.40  & 1.40 &  -0.28 & 1.16 \\
2   & 110 & $\pi/4$   & 450  &  -0.19  & 0.64 &  -0.13 & 0.54 \\
2   & 210 & 0         & 1050 &   0.15  & 0.50 &   0.10 & 0.42 \\ \hline
50  & 110 & $\pi/2$   & 800  &   0.96  & 3.29 &   0.66 & 2.74 \\
50  & 110 & $\pi/4$   & 150  &   0.96  & 3.29 &   0.66 & 2.74 \\
50  & 110 & 0         & 150  &   1.25  & 4.31 &   0.87 & 3.59 \\ \hline
\end{tabular}
\end{table}
\renewcommand{\arraystretch}{1}
\end{document}